\documentclass{jfm}
\usepackage{graphicx}
\usepackage{amsmath, mathtools}
\usepackage{epstopdf, epsfig}
\usepackage[per=frac, fraction=frac]{siunitx}
\usepackage{url}
\usepackage{comment}
\newcommand{\pdif}[2]{\ensuremath{\frac{\partial#1}{\partial#2}}}

\newcommand{\hatvec}[1]{\hat{\boldsymbol{\mathbf{#1}}}}
\renewcommand{\vec}[1]{\boldsymbol{\mathbf{#1}}}

\newcommand{\iu}{{i\mkern1mu}}

\newcommand\Ray{\mbox{\textit{Ra}}}

\usepackage{color}
\usepackage{soul}
\usepackage{float}

\shorttitle{Convection in dry salt lakes}
\shortauthor{J. Lasser, M. Ernst and L. Goehring}

\title{Stability and dynamics of convection in dry salt lakes}

\author{Jana Lasser\aff{*,1,2}
  \corresp{\email{lasser@csh.ac.at}}
Marcel Ernst\aff{*,1,3},
 \and Lucas Goehring\aff{4} \corresp{\email{lucas.goehring@ntu.ac.uk}}}

\affiliation{
\aff{*} Shared first author
\aff{1}Max-Planck Institute for Dynamics and Self-Organization, Am Fassberg 17, 37077 G\"ottingen, Germany
\aff{2} Complexity Science Hub Vienna, Josefst\"adterstrasse 39, 1080 Vienna, Austria
\aff{3} Department of Electrical Engineering and Computer Science, University of Kassel, Wilhelmsh\"oher Allee 71 - 73, 34121 Kassel, Germany
\aff{4} School of Science and Technology, Nottingham Trent University, Nottingham NG11 8NS, UK
}

\begin{document}

\maketitle

\begin{abstract} 
Dry lakes covered with a salt crust organised into beautifully patterned networks of narrow ridges are common in arid regions.  Here, we consider the initial instability and the ultimate fate of buoyancy-driven convection that could lead to such patterns. Specifically, we look at convection in a deep porous medium with a constant through-flow boundary condition on a horizontal surface, which resembles the situation found below an evaporating salt lake.  The system is scaled to have only one free parameter, the Rayleigh number, which characterises the relative driving force for convection.  We then solve the resulting linear stability problem for the onset of convection.  Further exploring the non-linear regime of this model with pseudo-spectral numerical methods, we demonstrate how the growth of small downwelling plumes is itself unstable to coarsening, as the system develops into a dynamic steady state.  In this mature state we show how the typical speeds and length-scales of the convective plumes scale with forcing conditions, and the Rayleigh number.  Interestingly, a robust length-scale emerges for the pattern wavelength, which is largely independent of the driving parameters.  Finally, we introduce a spatially inhomogeneous boundary condition---a modulated evaporation rate---to mimic any feedback between a growing salt crust and the evaporation over the dry salt lake. We show how this boundary condition can introduce phase-locking of the downwelling plumes below sites of low evaporation, such as at the ridges of salt polygons.
\end{abstract}

\begin{keywords}
Buoyancy-driven instability, Convection in porous media, Pattern formation, 
\end{keywords}

\section{Introduction}

\begin{figure}
    \centering
    \includegraphics[width=0.8\textwidth]{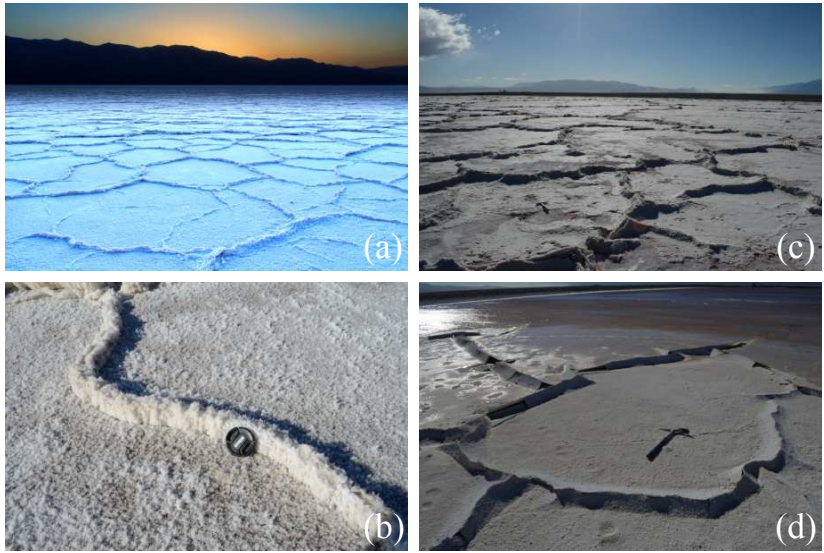}
    \caption{Typical salt polygon patterns at \textbf{(a,b)} Badwater Basin in Death Valley and \textbf{(c,d)} Owens Lake, both in California. (Image \textbf{(a)} courtesy~\citet{PHOTOGRAPHERSNATURE}, CC BY-SA 3.0.)}
    \label{fig:polygon_image}
\end{figure}

This study of porous-media convection is motivated by the patterns shown in figure~\ref{fig:polygon_image}.  The examples shown there are of dry salt lakes, or playa (\cite{BREIRE2000}), which are amongst the most inhospitable places on the surface of the earth. Dry lakes typically develop in arid environments, where evaporation outweighs precipitation and where mineral-rich groundwater is refreshed by inflow from surrounding regions of higher altitude (\cite{LOWENSTEIN1985,GILL1996,BREIRE2000}).   The heat fluxes through the surface of such salt deserts are important to understanding the water and energy balances in arid regions~\citep{BRYANT2002}.  Additionally, salt deserts are responsible for a significant part of the global emission of atmospheric dust~\citep{GILL1996, WASHINGTON2003, PROSPERO2002}.   However, despite the extreme conditions that prevail above ground, the water table of dry lakes often remains very near to the surface (\cite{GILL1996,BREIRE2000,BRYANT2003,NIELD2015}), allowing for active patterns of fluid flows within the pore spaces of the soil (\textit{e.g.}  \cite{WOODING1997a,VANDAM2009,STEVENS2009,TYLER1997}). As evaporation rates are high (\cite{TYLER1997,DEMEO2003,BRUNNER2004,GROENEVELD2010}), soluble salts accumulate in such regions, and precipitate into a solid salt crust covering the desert floor.  We have argued recently that these two processes, of subsurface flows and surface crust growth, are coupled together \citep{LASSER2019,LASSER_DISS2019}. In the main body of the present study we will focus on analysing the instabilities of the subsurface flow. Subsequently, we will discuss how this flow might support and interact with preferential precipitation of salt in certain areas on the surface.

Within the crust of a dry lake, captivating and beautiful patterns can emerge,  developing into a network of polygons, as shown in figure~\ref{fig:polygon_image}.  Around the world---from Owens Lake and Badwater Basin in California (\cite{LASSER2019,lasser2020surface}) or the Salt Lake desert of Utah \citep{CHRISTIANSEN1963} to the Great Salt Desert of Iran (\cite{KRINSLEY1970}), Salar de Uyuni in Bolivia and the Sua pan of Botswana (\cite{NIELD2015})---these salt polygons are usually expressed with a diameter of a couple meters, individually bounded by ridges a few centimetres high.  These regular patterns immediately draw the eye of an observer and the question of their origin arises.   The raised structures of polygonal ridge patterns in salt deserts also contribute to surface roughness~\citep{NIELD2015}. Under the effect of the often strong winds which blow over a desert's surface, dust is emitted from salt pans and is carried into the atmosphere. The surface roughness, alongside the salt chemistry and other crust characteristics, influences the uncertainty in modelling dust emission from desert landscapes~\citep{RAUPACH1993,MARTICORENA1995}.

\citet{CHRISTIANSEN1963} and \citet{KRINSLEY1970} attempted to explain the growth of salt polygons in dry lakes by the folding and cracking of the salt crust, respectively.  However, neither of these explanations is sufficient to explain the emerging polygonal shapes and, in particular, the robust length scale observed in nature around the world.   Specifically, both models would predict that the pattern wavelength is proportional to the thickness of the salt crust.  However, this wavelength is consistently 1-3 meters, in crusts ranging from sub-centimetre to several meters thick \citep{KRINSLEY1970,LOWENSTEIN1985,LOKIER2012,LASSER2019}.   Recently, buoyancy-driven convection, taking place within the wet porous sand below the salt crusts, was brought forward as an alternative candidate for a driving mechanism for pattern formation (\cite{LASSER2019}).  Here, we will model the onset of this mechanism as well as the maturation and scaling of the dynamics of buoyancy-driven convection that can occur below the surface of a dry salt lake.

\begin{figure}
    \centering
    \includegraphics[width=0.8\textwidth]{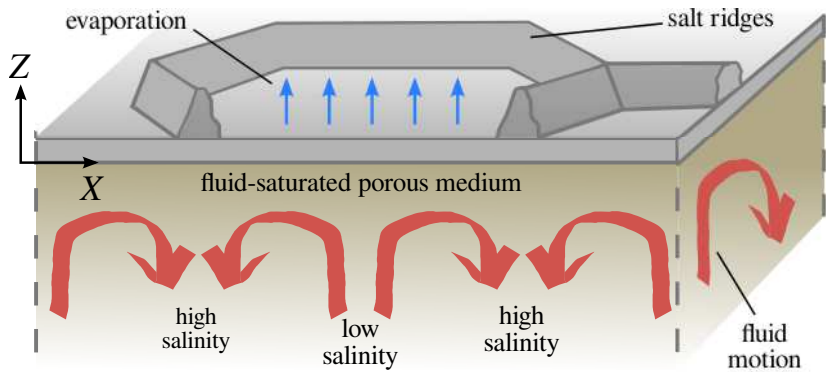}
    \caption{Sketch of the proposed buoyancy-driven flows in the soil beneath a polygon: water evaporates at the salt crust (arrows at surface) allowing the dissolved salt content to build up in the fluid-saturated porous medium below. When the salt-rich boundary layer becomes unstable, plumes of high salinity start sinking downwards below the salt ridges while fresh water rises in the middle of the polygon (red arrows).}
    \label{fig:model_sketch}
\end{figure}

Specifically, we will present a study of a model of solutal convection in a porous medium, as illustrated in figure~\ref{fig:model_sketch}. The porous medium can be interpreted as the sediment below the salt crust of a dry salt lake, and the solute as the salt dissolved in the groundwater that fills the porous medium. The salty groundwater is re-supplied by an influx of cleaner water from far below. Evaporation of water is enhanced both by wind and high temperatures, causing the precipitation of salt and growth of the crust at the surface. Due to the accumulation of salt near the surface the salinity, and thereby the density, of the water is higher there than it is further below. If the resulting density imbalance is large enough this configuration is unstable, leading to buoyancy-driven convective motion.

We model this problem in an idealised and simple 2D geometry, whose domain is deep compared to the dynamics that can arise near the surface. The domain has an upper boundary that is only permeable to fluid and which accounts for fluid loss through evaporation.  The lower boundary provides for the recharge of fluid from a reservoir of fixed salt concentration.  This model is investigated through a linear stability analysis, as well as with numerical simulations.  In the latter case we use periodic boundary conditions in the horizontal direction, and a lower boundary with a constant flux condition.  Both situations are given an initial condition in the form of a boundary layer of high salinity, formed just below the top boundary, in which salt diffusion is balanced by an upward advection of fluid. This setup is designed to mirror the conditions found below an evaporating salt lake, such as Owens Lake or Badwater Basin.  We investigate the onset of convective motion in the system at Rayleigh numbers $\Ray$ close to the critical value, $\Ray_\mathrm{c}$,  as well as the behaviour at higher Rayleigh numbers. Here, once the system is appropriately scaled, $\Ray$ is the only free parameter and can be interpreted as the dimensionless ratio between buoyancy forces and viscous dissipation. As the motivation for this work lies in the connection to pattern-forming processes in salt deserts, we also investigate the length-scales and time-scales of the resulting dynamics.  We also consider the effects of a modulated, non-uniform top boundary condition, which serves as a connection to surface feedback processes in nature.

More broadly, we note that the situation of buoyancy-driven convection occurring below an evaporating salt lake is also closely related to the convective overturning of CO$_2$,  dissolved in a porous medium filled with brine. This mechanism has become known for its importance to the sequestration of CO$_2$ in underground aquifers~(\textit{e.g.}  \cite{METZ2005,NEUFELD2010}; \cite{SLIM2010}; \cite{SLIM2013,SLIM,THOMAS2018,HEWITT2020}) to help mitigate the anthropogenic impact of CO$_2$ in the atmosphere. The similarity of models is such that an exchange of methods, insights and results is possible, in both directions.  In this context, our study relates to a variation on one-sided convection \citep{HEWITT2020b}.   For example, analogous to the study by~\citet{SLIM}, we numerically investigate the dependence of the dynamics of solutal convection in a porous medium on time, driven from one side, and we find similar regimes of behaviour, ranging from initiation to coarsening and eventually a dynamic steady state.  

The dynamics of thermally-driven porous media convection has also been extensively investigated in the past, for a variety of boundary conditions, and the equations are equivalent to the solutal-driven convection discussed here  (see \textit{e.g.} a recent review by \cite{HEWITT2020b}).  Of special interest has been the critical value of the Rayleigh number, $\Ray_\mathrm{c}$, above which a system is unstable to convective motion, for a variety of situations.  For example, a more well-studied case is where convection is driven from two sides, across a domain of fixed height but large width.  Here, when two perfectly conducting boundary conditions are chosen, similar to the setup of Rayleigh-B\'enard convection, then transport of heat across the system is purely conductive for $\Ray < 4\pi^2$~\citep{HORTON1945, LAPWOOD1948}. At higher $\Ray$, first steady and then perturbed convective rolls occur~\citep{BUSSE1972, GRAHAM1994}. For $\Ray \gtrsim 1300$, the dynamics enter a chaotic regime~\citep{OTERO2004, HEWITT2012}. 

There are also certain details already known about the onset of convection in the set of equations and one-sided boundary conditions that we study, which we will briefly summarise.   For a constant through-flow boundary condition---a situation that resembles a fluid-filled porous medium with surface evaporation---the onset of instability has been found to be at $\Ray_\mathrm{c}\approx14.35$~\citep{WOODING1960a, HOMSY1976, VANDUJIN2002} with a critical wavenumber of $k_\mathrm{c} \approx 0.76$ \citep{WOODING1960a}. Finite amplitude perturbations in the range of $8.59 < \Ray < 14.35$ were also observed to grow in simulations and experiments~\citep{VANDUJIN2002, WOODING1997a}.  In the slightly different case of a constant pressure boundary condition at the surface, a situation that resembles an evaporating salt lake covered with brine, the critical values of $\Ray_\mathrm{c} \approx 6.95$ and $k_\mathrm{c}\approx0.43$ are  smaller and thus the system is unstable for a greater range of parameters \citep{WOODING1960a}.  A linear stability analysis~\citep{WOODING1960a} as well as an energy minimisation method~\citep{VANDUJIN2002} were also used to calculate the neutral stability curves for both problems and to determine what range of wavenumbers are unstable at any given $\Ray$. In what follows we start by reproducing these theoretical results using an approach that develops particularly from the work of \cite{WOODING1960a}.  We are then able to complement the existing analysis by determining the most unstable mode for both constant pressure and constant through-flow boundary conditions, as well as solving for the growth rate of an arbitrary mode.  This is followed by a numerical study of the ultimate fate of this instability, including a discussion of how the resulting convection coarsens and scales in the highly non-linear regime.

\section{Governing Equations} \label{model}

The dynamics considered are those of fluid flow in the water-saturated porous soil of a dry salt lake and are described by the mass conservation of water and salt, as well as a detailed momentum balance.  The governing equations are those of incompressible flow, the mass conservation of salt with advection and diffusion and Darcy's law in the presence of gravity and are, respectively,
\begin{eqnarray}
\label{dim1}
\bnabla \boldsymbol{\cdot} \vec{q} &=& 0\,,\\
\label{dim2}
\varphi \pdif{S}{t} + \vec{q} \boldsymbol{\cdot} \bnabla S &=& \varphi\,D\,\bnabla^2 S\,,\\
\label{dim3}
\vec{q} &=& -\dfrac{\kappa}{\mu} \left(\bnabla p + \rho g \hatvec{z}  \right)\,.
\end{eqnarray}
These equations describe a fluid of superficial velocity, or flux, $\vec{q}$ and viscosity $\mu$ passing through a porous medium of porosity $\varphi$ and permeability $\kappa$ and carrying with it a dissolved salt of diffusivity $D$ and relative salinity $S$.  Flows are driven by a pressure $p$ and by buoyancy effects due to a gravitational acceleration, $-g\hatvec{z}$, and evolve over time $t$.  This system of equations has been studied in a wide variety of contexts, including geysers \citep{WOODING1960a}, analogues of Rayleigh-B\'enard convection (\textit{e.g.}  \cite{HORTON1945,LAPWOOD1948,ELDER1967,HEWITT2012}), solutal convection \citep{WOODING1997a,BOUFADEL1999} and carbon sequestration applications (\textit{e.g.}  \cite{LOODTS2014,HEWITT2014}).   For example, \cite{HEWITT2020b}
gives a recent review of their applications to vigorous convection in porous media.

Our aim is to determine the main length-scales and time-scales that emerge from this model of an evaporating salt lake.  The connection of the subsurface flow patterns to the surface crust can be seen in the form of Eq.~\ref{dim2}, from which the salinity flux into the crust follows as $E -\varphi D\partial_zS$, evaluated at $Z=0$.  We will return to the interactions of crust and flows in Section~\ref{sect:nonuniform}, after we describe how downwelling (low $\partial_zS$, so higher salinity flux into crust) and upwelling (high $\partial_zS$ and lower salinity flux) structures arise and scale in this system.

The relative salinity $S$ of the pore fluid depends on its density $\rho$ and the boundary conditions.  In our system fluid enters from below ($z\rightarrow -\infty$) at a background density $\rho_0$ and evaporates as a saturated solution, of density $\rho_1$, from the surface at $z = 0$. Between these limits
\begin{equation}
    \label{eq:salinity}
    S = \frac{\rho - \rho_0}{\rho_1 - \rho_0} = \frac{\rho - \rho_0}{\Delta\rho} 
\end{equation}
where $\Delta \rho = \rho_1 - \rho_0$. Thus, the salt-saturated fluid in contact with a solid salt crust has a relative salinity of $S = 1$, whereas the fluid feeding into the soil from some distant reservoir (which will still contain some dissolved salts) has $S=0$.

The formulation of Eqs. \eqref{dim1}-\eqref{eq:salinity} has a number of implicit assumptions that deserve mention.  First, it assumes a Boussinesq flow, such that density variations only affect the buoyancy term of Darcy's law.  It also neglects the effect of salt concentration on fluid viscosity: \citet{WOODING1960a}  has discussed this approximation, and shows how a more realistic salinity-dependant viscosity will produce a small stabilising effect; a similar result was obtained by \cite{BOUFADEL1999}, who showed that including a variable viscosity can slightly shift the balance of stability between competing modes of similar wavelengths.   Furthermore, the model treats the diffusivity $D$ as a constant, and thus ignores any velocity-induced dispersion (\textit{i.e.} Taylor dispersion, \citep{TAYLOR1953}).  \cite{WOODING1997a} show how to consider such effects, but for the field conditions measured at the salt playa of Owens Lake \citep{LASSER2019,lasser2020surface} dispersive effects should be negligibly small. Additionally, the equations treat the porosity and permeability of the soil as constant in space and time. For exemplary research on the types of effects that might be expected in systems with heterogeneous permeability or porosity, see~\citet{CHEN1998b, SHARP2009, LI2019, HEWITT2020}; and~\citet{HARFASH2013}.  Finally, it neglects thermal contributions to density changes, as being small compared to solutal effects.  This is justified by the relative magnitudes of these contributions: at Owens Lake, for example, we measured density changes due to salt content to be approximately 200 kg/m$^3$ \citep{CONCENTRATION, lasser2020surface}, whereas a 10$^\circ$C day-night temperature change would change the density of water by only about 1-2 kg/m$^3$. The buoyancy ratio, $N$, which gives the ratio of the solutal to thermal density differences, is thus of order $N\simeq100$. Effectively, this means that we neglect phenomena like double-diffusive convection, since the driving forces and typical speeds of such flows will be reduced by a factor of $1/N$, compared to solute-driven flows (see \textit{e.g.} \cite{Mojtabi2005} for a detailed review of such effects).

The non-dimensionalisation of our problem requires choosing a characteristic length $L$ and velocity (\textit{i.e.} flux) $\mathcal{V}$.  A natural time-scale then follows as $T = \varphi L/\mathcal{V}$.   Generally, for problems in porous media convection, scaling results in one of two clear choices for dimensionless groups, as Eqs. \eqref{dim2} and \eqref{dim3} become
\begin{eqnarray}
\label{dim2b}
\pdif{S}{\tau} + \vec{U} \boldsymbol{\cdot} \bnabla S &=& \frac{\varphi D}{L\mathcal{V}}\bnabla^2 S\,,\\
\label{dim3b}
\vec{U} &=& -\bnabla P - \frac{\kappa \Delta \rho g}{\mu \mathcal{V}} S \hatvec{Z}\,,
\end{eqnarray}
where $\vec{U} = \vec{q}/\mathcal{V}$, $\tau = t/T$, $Z = z/L$ and where $P$ is an appropriately rescaled pressure. For example, applications in carbon sequestration often use the limiting speed at which a dense parcel of fluid falls, $\mathcal{V}_\mathrm{B} = \kappa \Delta \rho g / \mu$, as a natural velocity scale (\textit{e.g.}  \cite{SLIM,HEWITT2014,HEWITT2020b}).  Similarly, in many cases a slab geometry of fixed height is modelled, where a scaling based on the layer thickness can be made \citep{RIAZ2003,RUITH2000,HEWITT2020b}.  We will instead follow a scheme where $\varphi D/L\mathcal{V} = 1$ and where the group in Eq.~\ref{dim3b} reduces to the Rayleigh number, $\Ray$.

For this, we focus on the situation of a solid salt crust through which there is a constant and uniform evaporation; the more complex case of a modulated evaporation rate will be explored in Section~\ref{sect:nonuniform}. As a boundary condition at the surface the upward fluid flux there must balance evaporation, such that the velocity component $q_z = E$ at $z=0$.   This will also give the average vertical flux of the pore fluid anywhere within the soil. Now, the governing equations allow for a simple stationary solution,
\begin{equation}
    S(z) = e^{zE/\varphi D}, \label{eq:steady-state-solution}
\end{equation}
which represents a dense boundary layer of fluid near the crust. For the one-dimensional problem, involving depth only, this is an attractive solution towards which transients will relax (see \textit{e.g.} \cite{WOODING1997a}). Following \cite{WOODING1960a} we therefore use $L = \varphi D / E$ as the characteristic thickness of the heavy boundary layer which can potentially develop.  This natural length scale is also the distance over which advective and diffusive effects will be of comparable magnitude.  The natural timescale $T = \varphi^2 D/E^2$ is then the time fluid takes to cross the boundary layer, in this stationary state, and the characteristic speed $\mathcal{V} = E$.  Applying these transformations results in the following non-dimensional formulation of Eqs. \eqref{dim1}-\eqref{dim3},
\begin{align}
  \vec{\nabla} \boldsymbol{\cdot} \vec{U} &= 0,\label{eq:incompressibilitynondim}\\
  \pdif{S}{\tau} + \vec{U} \boldsymbol{\cdot} \vec{\nabla} S &= \vec{\nabla}^2 S,\label{eq:saltconservationnondim} \\
  \vec{U} &=  -\vec{\nabla} P - \Ray\,S\,\widehat{\mathbf{Z}}\label{eq:darcynondim},
\end{align}
which is controlled by the Rayleigh number
\begin{align}
\Ray = \dfrac{\kappa\, \rmDelta\rho\, g}{\mu\,E}.
\label{eq:Rayleigh}
\end{align}
Here, $\Ray = \mathcal{V}_\mathrm{B}/E$ also represents the ratio of the characteristic speeds of a fluid parcel due to buoyancy and evaporation.  The dimensionless salinity flux into the surface, affecting crust growth, is simply $1-\partial_ZS$.  Finally, we note that the rescaled pressure, 
\begin{equation}
 P = \dfrac{\kappa}{\varphi\,\mu\,D} \left(p+\rho_0\,g\,z\right)\,, \nonumber
\end{equation} 
is now defined to include a contribution from the background fluid density.

In the rescaled system the boundary conditions are $U_Z = S = 1$ at $Z = 0$ (where $U_i$ are the components of $\vec{U}$), with the salinity approaching the limit of $S = 0$ at large depths.  These conditions are analogous to the thermal problem studied by \citet{WOODING1960a,HOMSY1976}.  They are appropriate to a dry salt lake as long as there is no significant ponding of surface water \citep{WOODING1997a, VANDUJIN2002}, although we will address the modifications to the model needed for ponding at the end of Section 3.1. The stationary solution has $\vec{U} = (0,0,1)$ everywhere, a salinity boundary layer given by $S = e^Z$ and a corresponding pressure $P = \Ray(1-e^Z)-Z$. In the following section we turn to investigate the stability of this solution, along with a time-dependent solution representing the growth of this boundary layer from an initially homogeneous lake.

\section{Linear Stability Analysis}
Here, we perform a linear stability analysis of our model for small perturbations around an initially unpatterned state. The methods used are inspired by those of \citet{WOODING1960a} and extend the range of his results to include an analytic series solution to the linear stability problem, along with predictions of the most unstable mode and first unstable mode.  For a base state we will focus on the stationary solution of a well-developed boundary layer, but will also consider the instabilities of a more general time-dependent solution. In section 4 we will confirm these results with a numerical implementation of our model of convection below a dry salt lake and further explore how they are modified at larger amplitudes, in other words in the nonlinear regime of the dynamics.

As set up in Section~\ref{model}, we consider an infinite half-space ($Z\leq 0$) of a three-dimensional porous medium saturated with saline water, which evaporates at the top boundary with a constant evaporation rate $E$ and which is recharged from below by a reservoir of constant salinity water. The methods used are inspired by those of Wooding (1960) and extend his results to include an analytic series solution to the problem.
To investigate the stability of a horizontially homogeneous base state, we add perturbations $\widetilde{\vec{U}}$, $\widetilde{S}$, $\widetilde{P}$ to its velocity, salinity and pressure fields, respectively.  The magnitudes of these perturbations are taken to be proportional to a small parameter, $\varepsilon_0$.  To leading order in $\varepsilon_0$ (specifically, ignoring the $\widetilde{\vec{U}}\boldsymbol{\cdot}\vec{\nabla}\widetilde{S}$ term, which is of order $\varepsilon_0^2$) the perturbations will grow or decay according to 
\begin{eqnarray}
\vec{\nabla} \boldsymbol{\cdot} \widetilde{\vec{U}} &= 0\,,
\label{eq:W7}\\
\pdif{\widetilde{S}}{\tau} + \pdif{\widetilde{S}}{Z} + \widetilde{U}_Z \,\pdif{S_0}{Z} - \vec{\nabla}^2 \widetilde{S} &= 0\,,
\label{eq:W9}\\
\widetilde{\vec{U}} + \vec{\nabla} \widetilde{P} + \Ray\,\widetilde{S}\,\widehat{\mathbf{Z}} &= \mathbf{0}\,.
\label{eq:W8}
\end{eqnarray}
Here, the only remaining term related to the base state is the base salinity $S_0 = S_0(Z,\tau)$, which arises from the non-linear aspects of the material derivative in Eq.~\eqref{eq:saltconservationnondim}.  The pressure term can be eliminated by taking the curl of Eq. \eqref{eq:W8}, as $\vec{\nabla}\times (\vec{\nabla}\widetilde{P})=0$.  By applying the curl twice, simplifying with Eq. \eqref{eq:W7} and considering the $Z$-component of the result, one finds that Eq. \eqref{eq:W8} implies that
\begin{equation}
  {\nabla}^2\,\widetilde{U}_Z + Ra\,\left(\p_X^2 + \p_Y^2\right) \,\widetilde{S} = 0.
  \label{eq:W10}
\end{equation} 

Solving for the dynamics of the perturbations requires providing for the appropriate boundary conditions.  In order to respect the original boundary conditions, $\widetilde{U}_Z = \widetilde{S} = 0$ at $Z = 0$.  As a second velocity condition, we assume a steady flow far from the unstable surface layer, such that the vertical perturbation to the velocity must decay to zero as $Z\rightarrow -\infty$.  Then, rearranging Eq.~\eqref{eq:W9} gives
\begin{equation}
  \widetilde{U}_Z = \partial S_0/\partial Z \left(\bnabla^2 \widetilde{S} - \pdif{\widetilde{S}}{Z} - \pdif{\widetilde{S}}{\tau}\right)\,,
  \label{eq:Wwm}
\end{equation}
which implies that 
\begin{equation}
\frac{1}{\partial S_0/\partial Z} \pdif{\widetilde{S}}{Z}\rightarrow 0 \quad \textrm{for} \quad Z\rightarrow-\infty.
  \label{eq:zinfty}
\end{equation}

Now, we can look at the evolution of the salinity perturbation, $\widetilde{S}$.  For this we follow the approach of \citet{pellew1940maintained} as well as \citet{WOODING1960a} and assume a separation of variables, such that
\begin{equation}
  \widetilde{S}(X,Y,Z,\tau) = F(Z) \Phi(X,Y) e^{\alpha \tau},
  \label{eq:Wpertubation}
\end{equation}
where $\Phi$ is a harmonic function satisfying $(\p_X^2 + \p_Y^2 + k^2)\Phi = 0$.  Here, $k$ is the characteristic wavenumber of the perturbation in the horizontal directions and $\alpha$ is its growth rate. For $\alpha > 0$ the amplitude of the perturbation increases and the system is unstable, whereas for $\alpha < 0$ the perturbation decays and the system is stable; the $\alpha = 0$ case will give the neutral stability curve.  Substituting Eq.~\eqref{eq:Wpertubation} into Eq.~\eqref{eq:W10} and using Eq.~\eqref{eq:Wwm} to simplify the result leads to the following eigenvalue equation for the height-dependant function $F(Z)$,   
\begin{equation}
  (\p_Z^2 - k^2)\left[\frac{1}{\partial S_0/\partial Z}\left(\p_Z^2 - \p_Z - k^2 - \alpha\right) F\right] = \Ray\,k^2\,F\,.
  \label{eq:W13}
\end{equation}
At this point, following the structure of Eqs.~\eqref{eq:Wwm} and \eqref{eq:W13}, we will also introduce
  \begin{equation}
  G(Z,\tau) \coloneqq \frac{1}{\partial S_0/\partial Z}\left(\p_Z^2 - \p_Z - k^2 - \alpha \right) F(Z)\,.
  \label{eq:W15b}
\end{equation}
At $Z = 0$ the boundary conditions of $F = G =  0$ follow from the fact that $\widetilde{U}_Z=\widetilde{S}=0$ there, and from Eq.~\eqref{eq:Wwm}.  In section~\ref{stat_sol} we will solve this problem for the stationary base state of $S_0 = e^Z$, whereas in section~\ref{dyn_sol} we will explore instabilities of the time-dependent case of a boundary layer developing from an initially homogeneous salinity field.

\subsection{Instabilities of the stationary base state} \label{stat_sol}

Using \texttt{DSolve} in \texttt{Mathematica} we obtained an analytical solution of the differential equation~\eqref{eq:W13} operating on $F(Z)$, for the stationary base state $S_0 = e^Z$. \citet{WOODING1960a} solved a similar problem for the special case of neutral stability ($\alpha = 0$), whereas we show a more general solution.  This solution is potentially a superposition of up to four independent infinite series.  Specifically, we first factor Eq.~\eqref{eq:W13} such that
\begin{equation*}
c_{0/1} = 1 \pm k,
\end{equation*}
\begin{equation*}
c_{2/3} = \frac{1\pm \sqrt{1 + 4 (k^2 + \alpha)}}{2} = \frac{1}{2} \left( 1 \pm \Psi\right)\,,
\end{equation*}
where $\Psi = \sqrt{1 + 4 (k^2 + \alpha)}$. The solutions then read
\begin{equation}
  F_{i}(Z) = \left(-k^2\, \Ray\,e^Z\right)^{c_i}\,H_i(Z)\quad \textrm{for }i \in \{1\ldots 4\}\,,
  \label{eq:W17}
\end{equation}
where $H_i(Z)$ is defined by
\begin{eqnarray}
  H_{0/1}(Z) &=& \,_0\mathrm{F}_3(;\{2c_{0/1}-1,c_{0/1}+c_2,c_{0/1}+c_3\};-k^2\,\Ray\, e^{Z})\,
  \nonumber \\
  H_{2/3}(Z) &=& \,_0\mathrm{F}_3(;\{2c_{2/3},c_{2/3}+k,c_{2/3}-k\};-k^2\,\Ray\, e^{Z})\,. \nonumber
\end{eqnarray}
These are hypergeometric functions of the form $\,_r \mathrm{F}_s$ with $r=0$ and $s=3$ \citep{KOEKOEK1998,askey2010generalized}.  They can be evaluated as series, for example, 
\begin{equation}
     H_{0}(Z) =\sum \limits_{n = 0}^{\infty}\left[ \frac{(- k^2\,\Ray\, e^{Z})^n}{n!\,(2c_0-1)_n(c_0+c_2)_n(c_0+c_3)_n} \right]\,, 
\end{equation}
where $(a)_n = a(a+1)(a+2)...(a+n-1)$ is the Pochhammer symbol for the rising factorial.  Since $r<s+1$ these series will always converge, except in the special cases where one of the terms in the denominator is $0$ \citep[p. 12]{KOEKOEK1998}; these exceptions occur when a term in the rising factorial sequence is $0$.  For example, $H_0$ is convergent except where $3/2 - \Psi/2+k=0,-1,-2...$, while $H_2$ will converge unless the same condition holds for $1/2 + \Psi/2-k$.  Hence, the solutions form a dense set in the $(\alpha,k)$ space.

From Eq.~\eqref{eq:zinfty} it follows that $F(Z)$ has to decay faster than $e^Z$ for $Z\rightarrow -\infty$ \citep{WOODING1960a}.  Therefore, of the four possible series described above, only those with $c_i>1$ are allowed, since $F_i \propto e^{Zc_i}$.  As $k\geq 0$, this condition eliminates the $c_1 = 1-k$ case, whereas $c_0 = 1+k$ always leads to a valid solution.  Similarly, as long as $k^2+\alpha>0$, then $c_3<1$ and $c_2>1$.   Thus, a general solution under these conditions (which include all unstable cases) can be given by $F(Z) = C_0 F_0(Z) + C_2 F_2(Z)$, for some real constants $C_0$ and $C_2$.  

\begin{figure}
\centering
  \includegraphics[width=\textwidth]{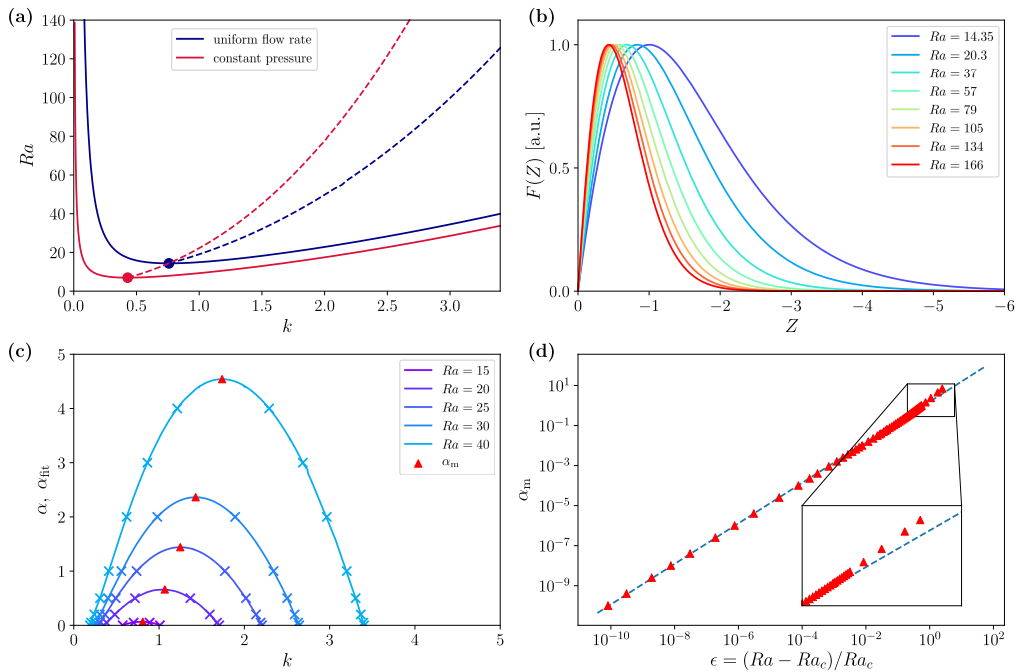}
  \caption[Neutral stability curves and most unstable wavenumber.]{Linear stability analysis and behaviour of numerical simulations near the critical point.  \textbf{(a)} Neutral stability curves (solid lines), critical points at $\Ray_\mathrm{c}$, $k_\mathrm{c}$ (filled circles) and the most unstable mode (dashed lines), according to the linear stability analysis for a constant salinity at the upper surface and either uniform vertical flow (blue) or constant pressure (red) there. For all subsequent panels, only the uniform flow problem is considered. \textbf{(b)} Normalised eigenfunctions, $F(Z)$, for the most unstable mode, calculated at various $\Ray$.  \textbf{(c)} Theoretical prediction of growth rates $\alpha$ as a function of wavenumber $k$  for a selection of $\Ray$ close to the onset of instability (solid lines). Their maximum values, $\alpha_\mathrm{m}$, are indicated as red triangles. Measurements of the growth rates of single-mode perturbations ($\alpha_{fit}$) in the corresponding simulations are indicated as crosses, as discussed in the appendix (see figure~\ref{fig:amplitude_measurement}). \textbf{(d)} Theoretical values for the growth rate of the most unstable mode, $\alpha_\mathrm{m}$, for different conditions.  Near the critical point, where $\epsilon = (\Ray - \Ray_\mathrm{c})/\Ray_\mathrm{c}$, the growth rate follows a power-law scaling, $\alpha_{\mathrm{m}} = \epsilon^\gamma$, but begins to deviate from $\gamma = 1$ when $\epsilon \gg 0$ (see inset). }
  \label{fig:neutral_stability_and_growth_rates}
\end{figure}

For our model, as mentioned earlier, the top boundary conditions are satisfied if and only if $F(0) = G(0) = 0$.  Using Eq.~\eqref{eq:W15b} this condition can be written as \mbox{$F(0) = C_0\,F_0(0) + C_2\,F_2(0) = 0$} and $G(0) = C_0\,G_0(0) + C_2\,G_2(0) = 0$.  Consequently, we know that either $C_0 = C_2 = 0$ or
\begin{equation}
\left|\begin{matrix}
F_0(Z) & F_2(Z) \\ G_0(Z) & G_2(Z)
\end{matrix}\right|_{Z=0} = F_0(0) \,G_2(0) - F_2(0)\,G_0(0)  = 0\,. \label{eq:Det}
\end{equation}
Since only the non-trivial solution is physically relevant, applying this constraint gives a relationship between $\Ray$, $k$ and $\alpha$ and allows one of these parameters to be determined by fixing the other two.   For example, as in  figure~\ref{fig:neutral_stability_and_growth_rates} \textbf{(a)}, we can use Newton's method to find the roots of the determinant given in Eq.~\eqref{eq:Det} for some particular values of $k$ and $\alpha$, and hence find the smallest $\Ray>0$ satisfying the boundary conditions (\textit{e.g.} for the neutral stability curve, $\alpha = 0$).  Finally, we note that \citet{WOODING1960a} applied slightly  different upper boundary conditions to his model, which here would correspond to constant salinity, $S=1$, but with a fixed surface \textit{pressure} instead of a constant surface through-flow.   This could simulate a shallow layer of salt-saturated water at the surface, for example, or a water-logged crust.  For the fixed pressure case there would be no horizontal flows along the surface (\textit{i.e.} it would act as a no-slip boundary), and the incompressibility condition (Eq.~\eqref{eq:W7}) then implies that $\partial_Z \widetilde{U}_Z=0$ there.  For that scenario the arguments and solutions presented above remain valid, but require the modified boundary conditions of $F = \partial_Z\,G = 0$ at $Z = 0$.   The determinant in Eq.~\eqref{eq:Det} is similarly revised, to read $F_0(0) \,\partial_Z\,G_2(0) - F_2(0)\,\partial_Z\,G_0(0)  = 0$.

\begin{table}
  \begin{center}
\def~{\hphantom{0}}
   \begin{tabular}[h]{l|cc}
  Boundary condition & $\Ray_\mathrm{c}$ & $k_\mathrm{c}$ \\ \hline
  Uniform flow rate & $14.35$ & $0.7585$ \\
  Constant pressure & $6.954$ & $0.429$
  \end{tabular}
  \caption{Critical Rayleigh number, $\Ray_\mathrm{c}$, and wavenumber, $k_\mathrm{c}$, from linear stability analysis.}
  \label{tab:critical}
    \end{center}
\end{table}

Some results from the solutions to our linear stability problem are given in figure~\ref{fig:neutral_stability_and_growth_rates} and table~\ref{tab:critical}.  In figure~\ref{fig:neutral_stability_and_growth_rates} \textbf{(a)} we show the neutral stability curve and most unstable mode for both types of boundary conditions: results for constant evaporative flux are shown in blue and constant pressure in red.  We also indicate the critical points for both cases and details of the critical Rayleigh number, $\Ray_\mathrm{c}$, and its corresponding critical wavenumber, $k_\mathrm{c}$, are given numerically in table~\ref{tab:critical}.  For the constant pressure case the neutral stability curve and critical point are consistent with the results of \citet{WOODING1960a}. For the constant flux case, the corresponding results are consistent with \citet{HOMSY1976} and \citet{VANDUJIN2002}.  The most unstable mode calculations provide additional predictions, which we will use to validate our numerical model of convection.  

The eigenfunctions, $F(Z)$, corresponding to the most unstable modes of scenarios of different $\Ray$ are shown in figure~\ref{fig:neutral_stability_and_growth_rates} \textbf{(b)}.  These have been normalised to have a maximum value of 1.  In all cases the general shape of $F(Z)$ is similar to the solution sketched by \cite{WOODING1960a} for constant pressure conditions and at the critical point.  Furthermore, the eigenmodes do not undergo any significant qualitative changes as $\Ray$ increases, but rather the peak gradually narrows and shifts to shallower depths, reflecting the $k$-dependence of Eq.~\eqref{eq:W17}.

In figure~\ref{fig:neutral_stability_and_growth_rates} \textbf{(c)} we show the growth rate $\alpha$ for various modes and conditions just above the critical point  (alongside corresponding results from our numerical model, for purposes of validation).  These are consistent with a type-I (finite wavenumber, see \cite{CROSS2009}) instability. In the following section we will confirm these results with a numerical implementation of our model of convection below a dry salt lake and further explore how they are modified at larger amplitudes, in other words in the nonlinear regime of the dynamics. 

Finally, and introducing $\epsilon = (\Ray - \Ray_\mathrm{c})/\Ray_\mathrm{c}$ to describe the proximity of the system to the critical point, figure~\ref{fig:neutral_stability_and_growth_rates} \textbf{(d)} shows that the growth rate of the most unstable mode, $\alpha_{\mathrm{m}}$, scales linearly with $\epsilon$ near the critical point, as expected for a type-I instability \citep{CROSS2009}.  Given this relationship, when we make quantitative comparisons of \textit{e.g.} velocities at different $\Ray$ in what follows, we will often find it convenient to re-scale time, and thus define $\widehat{\tau} = \tau\epsilon$.

\subsection{Instabilities of a transitory base state} \label{dyn_sol}

\begin{figure}
    \centering
    \includegraphics[width=\textwidth]{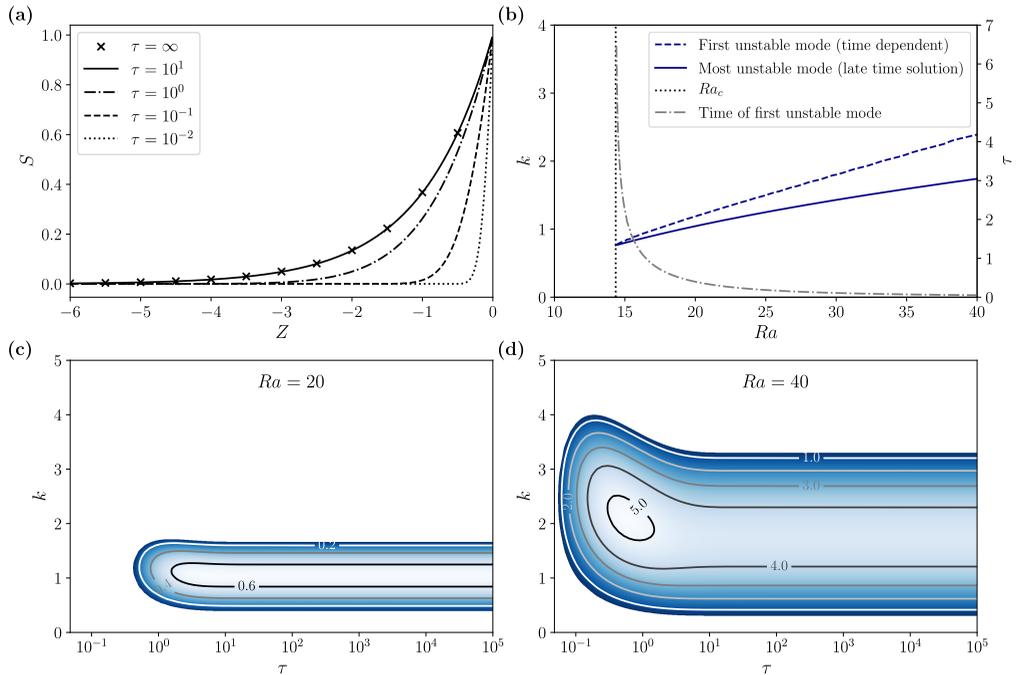}
    \caption{Linear stability analysis of time-dependent solution. \textbf{(a)} For the case of an evaporating salt lake with an initially constant salinity everywhere ($S=0$), and where the surface conditions are changed to $S=1$ at $\tau=0$, a time-dependent boundary layer will develop (dashed lines).  This layer will relax to the stationary base state $e^Z$ over a timescale $\tau\sim 1$. \textbf{(b)} The linear stability of this boundary layer was evaluated to determine the first moment of instability (right y-axis), along with the first unstable mode (left y-axis).  The most unstable mode of the stationary base state (from figure~\ref{fig:neutral_stability_and_growth_rates} \textbf{(c)}, corresponding to late-time solution) is included for comparison. For \textbf{(c)} $\Ray=20$ and \textbf{(d)} $\Ray=40$ we show how the growth rates of the spectrum of unstable modes depend on time.}
    \label{fig:time_dependent_perturbation}
\end{figure}

Before proceeding to a full numerical simulation of the salt playa problem, we will consider how instabilities would arise for the case of a transient initial condition. This analysis is inspired by that of \cite{SLIM2010}, who presented a time-dependent linear stability analysis of the related problem of convection without through-flow, but including the case of a permeable upper surface.  Specifically, we will consider the situation where the initial salinity everywhere is equal to that of the reservoir, $S=0$, and where the $S=1$ boundary condition is suddenly applied to the surface at $\tau=0$. This could describe the situation of a rapid change in conditions, such as the abrupt flooding of a surface by brine, for example, or the rise of a buried water table to the surface, which could reactivate crust growth.

For this initial condition, the system has the transient solution \citep{WOODING1997a}
\begin{equation}\label{eq:eig}
    S_0(Z,\tau) = e^{Z/2}\left[\frac{1}{2}e^{Z/2} \mathrm{erfc}\left(\frac{-Z-\tau}{2\sqrt{\tau}}\right) + \frac{1}{2}e^{-Z/2} \mathrm{erfc}\left(\frac{-Z+\tau}{2\sqrt{\tau}}\right) \right].
\end{equation}
As shown in figure~\ref{fig:time_dependent_perturbation} \textbf{(a)}, this solution relaxes to the stationary base state, $e^Z$, over a timescale of $\tau \sim 1$.  The question now is whether the instabilities that can occur during such a transient phase will be consistent with those of the stationary base state, or whether they will allow for additional behaviours.  We will show that the potential instabilities during the diffusive growth of the boundary layer are broadly similar to the instabilities of the well-developed boundary layer given in section 3.1.

Equation~\ref{eq:W13} can be rearranged into an eigenvalue equation of the form $\mathsfbi{A}F = \alpha \mathsfbi{B}F$:
\begin{equation}
\left((\p_Z^2 - k^2)\left[\frac{1}{\partial S_0/\partial Z}\left(\p_Z^2 - \p_Z - k^2\right) \right]-\Ray k^2\right)F = \alpha\,(\p_Z^2 - k^2)\, \left[\frac{1}{\partial S_0/\partial Z}\right]F\,.
  \label{eq:eig2}
\end{equation}
The largest eigenvalue here gives the growth rate of the most unstable perturbation for any particular mode $k$ at any instant $\tau$ and Rayleigh number $\Ray$.   We solved this eigenvalue problem numerically, using a Chebyshev differentiation matrix \citep{TREFETHEN2000} for the differential operators, $\partial_Z$, acting on $F$ and $S_0$, the base state given in Eq.~\ref{eq:eig}, and on a domain of a finite height $h$.  The lower boundary has a constant flux $U_Z = 1$ of water with relative salinity $S=0$, as will be used throughout Section 4.  The late-time ($\tau = 100$) solutions were used to validate the method against the results shown in figure~\ref{fig:neutral_stability_and_growth_rates} \textbf{(b,c)}, and agreed well for domains with a height of at least $h\simeq 5$.  Since large $h$ can introduce issues with numerical precision, due to the very rapid decay of Eq.~\ref{eq:eig} with depth, we therefore used $h=7.5$ in what follows.  

For the eigenvalue analysis of the transient base state we used Newton's method to search for the time that the first mode went unstable, for any given $\Ray$.  In figure~\ref{fig:time_dependent_perturbation} \textbf{(b)} we show how the time of the first instability, and the value of the first unstable mode, depend on $\Ray$.  For comparison, we also include the most unstable mode of the stationary base state, from figure~\ref{fig:neutral_stability_and_growth_rates} \textbf{(a)}.  The first unstable mode is similar to the most unstable mode, but consistently slightly higher.  The instability sets in rapidly, suggesting that for these initial conditions there will be competition between the growth of the boundary layer, and the growth of the instability.  In section 4.4 we will explore this competition further, using fully non-linear numerical simulations.

To investigate how the the spectrum of unstable modes evolves through time, as the boundary layer fills up, we also calculated the growth rate for a grid of different $\tau$ and $k$.  Examples are shown in figure~\ref{fig:time_dependent_perturbation} \textbf{(c,d)} for the cases of $\Ray=20$ and 40, respectively.  We find that the range of unstable modes does not change dramatically over time.  Rather, once a mode becomes unstable, it generally remains so.  For example, for $\Ray = 40$, modes between $k=0.212$ and $3.412$ are unstable at $\tau = 10$, and this range is effectively coincident with the range of unstable modes as $\tau \to\infty$ (namely, from $k= 0.211$ to $3.406)$.  The exception to this behaviour is a range of wavenumbers at the highest end of the unstable spectrum.  Continuing our example for $\Ray = 40$, the wavenumbers from $k = 3.41$ to $4.09$ are stable to small perturbations at long times, but unstable for some period of the transient.  Note that this behaviour is different from the case of the time-dependent diffusion into a finite porous layer \textit{without} through-flow (for either an impermeable or permeable surface, \cite{SLIM2010}), where all modes eventually return to a stable situation.

Finally, we note that a more comprehensive analysis of the time-dependent stability problem could be made by non-modal stability theory, as has been done for the related problem of solutal convection without through-flow (see \textit{e.g.} \cite{RAPAKA2008,SLIM2010}).  Alternatively, we will return to present numerical simulations with a time-dependent base state in Section 4.4.  What we can conclude here, however, is that the range of unstable modes does not change significantly throughout a transient phase, and that the first unstable mode for the time-dependent case is close to the most unstable mode found in section~\ref{stat_sol}, for the stationary base state.

\section{Scaling relationships in the nonlinear regime}

For our model of subsurface convection in a dry salt lake the linear stability analysis predicts that a salt-rich boundary layer is unstable to convective rolls above a Rayleigh number of about 10, depending on the exact nature of the boundary conditions.  Now we focus on the longer-time behaviour of the model and on the scaling of the convective dynamics as the system matures towards a dynamic steady state.  To this end, we numerically solved the governing equations, \eqref{eq:incompressibilitynondim} to \eqref{eq:darcynondim}, on rectangular domains with periodic boundary conditions in the horizontal direction.  The deep aquifer is modelled as a lower boundary at a fixed background salt concentration, such that $S=0$, and a constant recharge rate, $U_Z = 1$.  As an initial condition, we focus on the $S = e^Z$ time-independent solution of a well-developed diffusive boundary layer; different initial conditions will be explored in section 4.4. Details of the implementation of the numerical simulation, which follow the pseudo-spectral approach of~\citet{RIAZ2003,RUITH2000,CHEN1998a}, are given in appendix A and the code itself is available on a public repository \citep{simulationcode}.   

The simulations were first validated against the theoretical predictions of growth rates.  Specifically, for $\Ray$ between 15 and 40 we added small perturbations of a single wavenumber to the initial conditions.   The growth rate of this mode, $\alpha_{fit}(k,\Ray)$, was measured by a fit to half the peak-to-peak amplitude of the perturbation in the linear growth regime (see validation section and figure~\ref{fig:amplitude_measurement} in the appendix). As shown by the crosses in figure~\ref{fig:neutral_stability_and_growth_rates} \textbf{(c)}, the simulated $\alpha_{fit}$ agree with the results of the linear stability analysis.    

\begin{figure}
    \centering
    \includegraphics[width=\textwidth]{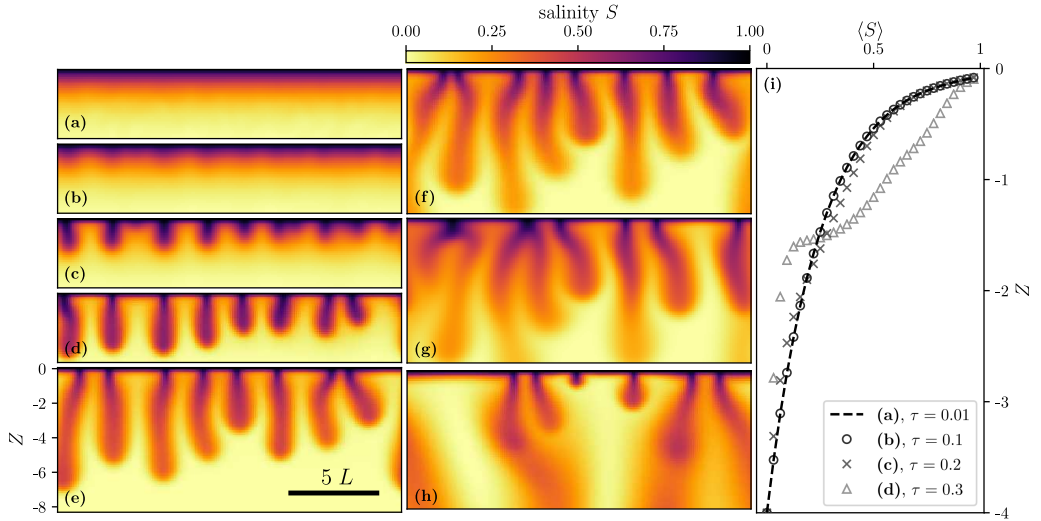}
    \caption{Snapshots of the relative salinity $S$ at different times $\tau$ for a simulation with $\Ray = 100$.  Panels show \textbf{(a)} the initial salinity distribution at $\tau = 0.01$, the linear growth regime at \textbf{(b)} $\tau = 0.1$ and \textbf{(c)} $\tau = 0.2$, the flux growth regime at \textbf{(d)} $\tau = 0.3$ and \textbf{(e)} $\tau = 0.5$, the merging regime at \textbf{(f)} $\tau = 0.75$ and \textbf{(g)} $\tau = 1$ and the re-initiation regime at \textbf{(h)} $\tau = 3$.  Panel \textbf{(i)} shows the horizontally averaged salinity distributions for the snapshots \textbf{(a)} to \textbf{(d)}. The domain size of the simulation was $40\,L\times 40\,L$, and only part of the domain is shown in each case.   The scale is the same for all snapshots, and given in panel \textbf{(e)}.}
    \label{fig:simulation_snapshots}
\end{figure}

More generally, we added low levels of random noise, at all $k$, to the initial conditions.  We then performed simulations for a selection of Rayleigh numbers ranging from 15 to 2000. The spatial resolution of the simulations increased with increasing $\Ray$, so as to be able to resolve all important features of the dynamics.  Consequently, the system sizes were adjusted (smaller width $W$ and height $H$ for higher $\Ray$) to keep the computational cost of the simulations manageable. Spatial resolutions and domain dimensions are listed in the appendix, table~\ref{tab:simulation_stats}. Furthermore, in what follows, all uncertainty ranges given represent the standard deviations of properties measured in ensembles of 5 to 10 runs for each set of conditions: they are intended to showcase the variability seen between simulations.  Snapshots of an example simulation at $\Ray = 100$ and at different times $\tau$ are displayed in figure~\ref{fig:simulation_snapshots}.  Supplementary movies S1, S2 and S3 also give the results of three example simulations at $\Ray = 30$, $\Ray=100$ and $\Ray=1000$, respectively. 

As the simulations proceed they pass through several distinct regimes of dynamics, which can be related to those given by~\citet{SLIM} for a similar problem motivated by one-sided convection beneath a CO$_2$ pool (in other words, without the evaporative flux of our model). At early times, as shown in figure~\ref{fig:simulation_snapshots} \textbf{(b,c)}, we observe a regime of the \textit{linear growth} of high salinity plumes, at a wavelength corresponding to the most unstable mode of the linear instability. This is followed by a \textit{flux-growth} regime, where the downwelling plumes strip the boundary layer of its heavy burden of solute.  Such a thinning of the boundary layer can be seen in figure~\ref{fig:simulation_snapshots} \textbf{(d,e)}.  The next regime is the \textit{merging} regime, during which time the plumes begin to influence each other \textit{via} long-range interactions in the horizontal velocity field. As a result, nearby plumes are attracted to each other and merge together to form larger plumes, as is happening in figure~\ref{fig:simulation_snapshots} \textbf{(f,g)}. Once enough plumes have merged, the high salinity boundary layer feeding the plumes begins to grow again, and it thickens until small proto-plumes start emerging at the top boundary and we enter a \textit{re-initiation} regime, as shown in figure~\ref{fig:simulation_snapshots} \textbf{(h)}. These proto-plumes are typically attracted to and then swept into the larger pre-existing plumes.   After this time the system settles into a long dynamic steady state period, which lasts until the deepest plumes start interacting with the lower boundary, after which point we typically stop the simulation.

It is worth noting that these simulations start with a well-developed boundary layer, in which upwards advection and diffusion back down the concentration gradient balance.  As such, they do not show a clearly-defined diffusive regime, which would correspond to the initial growth of this layer, up to the first moment of instability, starting from a homogeneous solute distribution as initial condition (\textit{i.e.} the dynamics shown in  figure~\ref{fig:time_dependent_perturbation} \textbf{(a)}).  Instead, at early times the boundary layer is deformed by the growth of the unstable modes.  To illustrate this, in figure~\ref{fig:simulation_snapshots} \textbf{(i)} we show the horizontally-averaged salinity distributions for the simulation snapshots displayed in panels \textbf{(a)} to \textbf{(d)}.  A diffusive regime, analogous to the first regime reported by \cite{SLIM}, would be expected for a homogeneous initial condition.  Indeed, for such an initial condition, the time of the first unstable mode, reported in figure~\ref{fig:time_dependent_perturbation} \textbf{(b)} shows how the duration of the diffusive regime would depend on $\Ray$.  Similarly, we see a diffusive regime in simulations (see section 4.4) started with a less-developed boundary layer.    

We also do not observe a clear \textit{shut-down} regime, since our boundary conditions allow for a through-flow of solute (rather than a constant build-up of salt that would be seen for impermeable boundary conditions).  Indeed, we argue that the variations in salt flux to the surface, caused by the presence of the plumes, are important for the surface patterning seen in dry salt lakes \citep{LASSER2019}.

\subsection{Scaling of the plume velocity}
The convective dynamics seen in the simulations tend to become more vigorous at higher Rayleigh numbers.  This reflects the interpretation of $\Ray$ as the ratio of the natural speeds of flows driven by buoyancy and evaporation.  To quantify this relationship we measured the maximum speed of the plumes relative to the background flow by calculating $V(\tau) = \mathrm{max}(|U_Z - 1|)$ over the entire numerical domain at each time step of various simulations.  

\begin{figure}
    \centering
    \includegraphics[width=\textwidth]{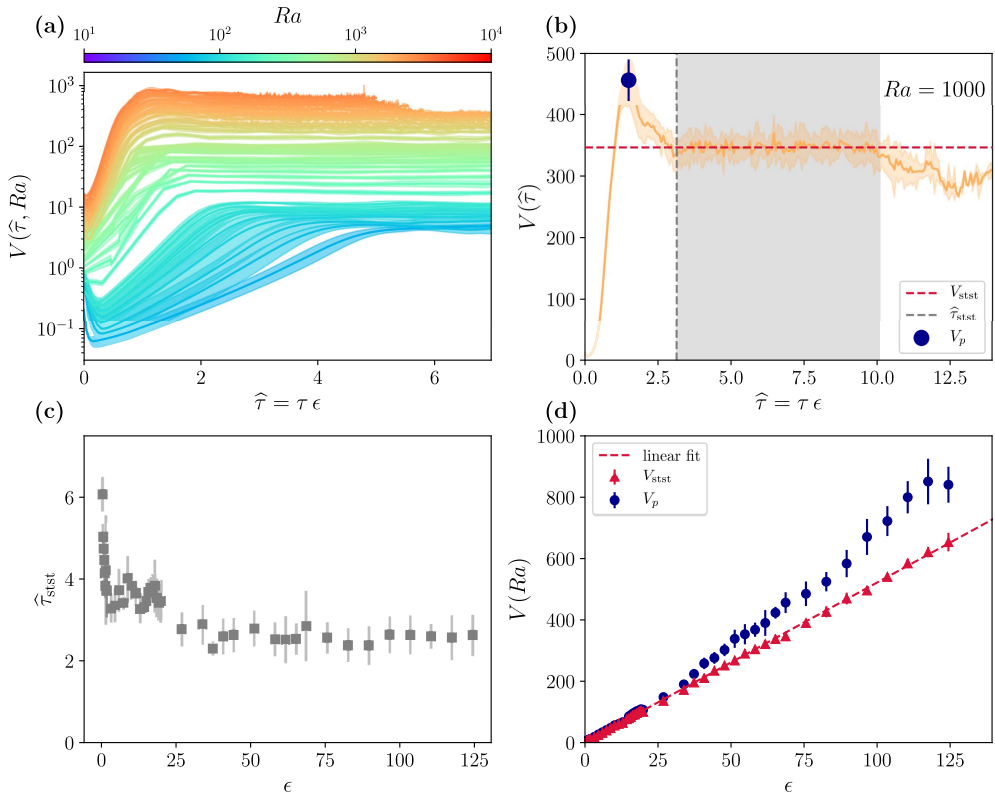}
    \caption{Development of the maximum plume speed, $V$, for different driving conditions and where $\epsilon = (\Ray - \Ray_\mathrm{c}) / \Ray_\mathrm{c}$. \textbf{(a)} Evolution of the plume speeds in simulations at different $\Ray$ show an initial transient, followed by a peak in velocity as the first plumes grow and merge and a subsequent steady state condition. \textbf{(b)} For the case of $\Ray = 1000$ we show results from a single simulation as a solid line, while the standard deviation within the corresponding ensemble is given by the shaded area. The blue dot, horizontal red dashed line, vertical grey dashed line and the grey rectangle indicate the measured peak speed $V_p$, the dynamic steady state speed $V_\mathrm{stst}$, the beginning of the dynamic steady state $\widehat{\tau}_\mathrm{stst}$ and the time period used to calculate $V_\mathrm{stst}$, respectively. \textbf{(c)} The time taken until the system reaches a steady state, $\widehat\tau_\mathrm{stst} = \tau_\mathrm{stst}\epsilon$, settles down to a constant value of $\widehat\tau_{\mathrm{stst}}\approx2.5$ for large enough $\Ray$ and $\epsilon$. \textbf{(d)} Development of $V_\mathrm{p}$ and $V_\mathrm{stst}$ with $\epsilon$, showing how the plume velocity in the steady state scales linearly with the proximity to the critical conditions, $\epsilon$, whereas the initial overshoot of the plume speeds becomes more apparent at larger $\Ray$. }
    \label{fig:velocity_scaling}
\end{figure}

In figure~\ref{fig:velocity_scaling} \textbf{(a)} the time development of $V$ is given for a range of Rayleigh numbers. There are initially small fluctuations in this speed, as the dominant unstable modes are selected from our broad-spectrum perturbation.   After this brief initial transient the plume speed, characterising convection, increases until it reaches a peak and then plateaus at an essentially constant value.  For $\Ray > 100$ the initial peak can overshoot its plateau value significantly, as can be seen in figure~\ref{fig:velocity_scaling} \textbf{(b)} for a simulation at $\Ray=1000$.  In all these simulations the initial variations in $V$ correspond to the linear growth, flux-growth and merging regimes whereas the plateau corresponds to the re-initiation regime, which behaves as a dynamic steady state. Again, if we instead used homogeneous initial conditions we would also expect an initial diffusive regime to precede the linear growth phase (i.e. up to the time of first instability given in figure 4 \textbf{(b)}).  At very long times the plumes start to interact with the lower boundary and the convection starts to weaken, hence $V$ decreases.  

For every simulation we identified the peak speed, $V_\mathrm{p} = \mathrm{max}(V(\tau))$, as shown for example by the blue dot in figure~\ref{fig:velocity_scaling} \textbf{(b)}. We also estimated the beginning of the subsequent dynamic steady state, $\tau_\mathrm{stst}$, as the time after which $\partial V/\partial \tau < 0.1$.  This is found to be, empirically, a good measure of the time (\textit{e.g.} grey dashed line in figure~\ref{fig:velocity_scaling} \textbf{(b)}) after which the relative plume speed is no longer systematically changing, although it continues to fluctuate randomly after this point. As shown in figure~\ref{fig:velocity_scaling} \textbf{(c)}, this time to develop a steady state approaches a value of $\tau_\mathrm{stst} \approx 2.5/\epsilon$ (or, alternatively, $\widehat\tau_\mathrm{stst} \approx 2.5$), for large enough $\epsilon$. This can be explained by the time the perturbations need to grow, since the initial amplitude of the perturbations to the salinity field is independent of $\Ray$, whereas the growth rate of the most unstable mode is proportional to $\epsilon$.  Although the time needed for the disturbances to saturate will depend on the initial amplitude of these perturbations, the scaling of $\tau_{\mathrm{stst}}$ demonstrates that $\epsilon$ remains a useful characterisation of the relative vigour of the convective process well into the non-liner regimes.  

To emphasise the scaling of the system times and speeds with $\epsilon$, we also calculated a steady state speed, $V_\mathrm{stst}$, as the average of $V$ in the time window between $\tau_\mathrm{stst}$ and $\tau_\mathrm{stst} + 7 \epsilon$.  This window was chosen to be as wide as possible, so as to provide a stable average value, while still avoiding the start of the shut-down regime for the largest $\Ray$ in our study, where the shut-down of convection also approaches fastest.  For example, for the $\Ray=1000$ case the this time-window and $\V_\mathrm{stst}$ are indicated as the grey shaded area and the red dashed line in figure~\ref{fig:velocity_scaling} \textbf{(b)}, respectively.  In figure~\ref{fig:velocity_scaling} \textbf{(d)} we show how both $V_\mathrm{p}$ and $V_\mathrm{stst}$ vary with $\epsilon$, and hence $\Ray$.  This shows that the characteristic plume speed in the steady state, \textit{i.e.} the re-initiation regime, increases linearly with $\epsilon$.  It also shows how the overshoot of the plume speeds becomes more significant at larger $\Ray$, when the system starts out in a more unstable initial configuration.  

Finally, we note that the use of a maximum speed to characterise the rate of convection has the potential to lead to overestimates.  As such, we confirmed the scaling of the plume speed by considering the average speed of all downwelling plumes at a depth of $Z = -1$ and at the moment where the first plume tip reached a depth of $Z=-2$ (specifically, when $S=0.5$ was first exceeded there).  The results of this measurement are consistent with those otherwise presented here, in that the average plume speed scales linearly with $\epsilon$~\citep{ernst2017numerical}.  Similarly computing the average \textit{upwelling} speed at this time and depth gives the same scaling.  Thus, for a number of different metrics, the plume velocity in the system is an indicator of how fast the dynamics pass through the different regimes, and shows a simple linear scaling with $\epsilon$.  This can be readily understood, given that $\Ray= \mathcal{V}_\mathrm{B}/E$ represents the ratio of the characteristic speeds of buoyancy and evaporation.

\subsection{Scaling of the plume wavelength}

As the buoyancy-driven convection in our model evolves away from its initial instability, the number and spacing of the salt plumes can change.  Already, in figure~\ref{fig:simulation_snapshots}, we showed that the pattern has a tendency to coarsen during the merging regime, before a steady state develops where plume merging and re-initiation balance each other.  Thus, the most unstable mode calculated by the linear stability analysis is unlikely to be representative of the long-term pattern that would be seen for examples of this convective process occurring in nature.  In the following, we will quantify this coarsening behaviour with the aim of determining the spatial scale of convection expected in systems that are in a dynamic steady state, and long after their onset in time. 

To characterise the spatial structure of the dynamics in a simulation at any given depth $Z$ and time $\widehat\tau = \tau \,\epsilon$, we measured the effective wavelength $\Lambda=W / N$ of the plumes, where $W$ is the width of the simulated domain and $N$ is the number of downwelling plumes of high salinity. For this, the positions of the plumes are identified as the maxima in the salinity along a horizontal profile of depth $Z$. We then take \mbox{$k = 2\pi /\Lambda$} to be the corresponding wavenumber of the plumes. 

\begin{figure}
    \centering
    \includegraphics[width=\textwidth]{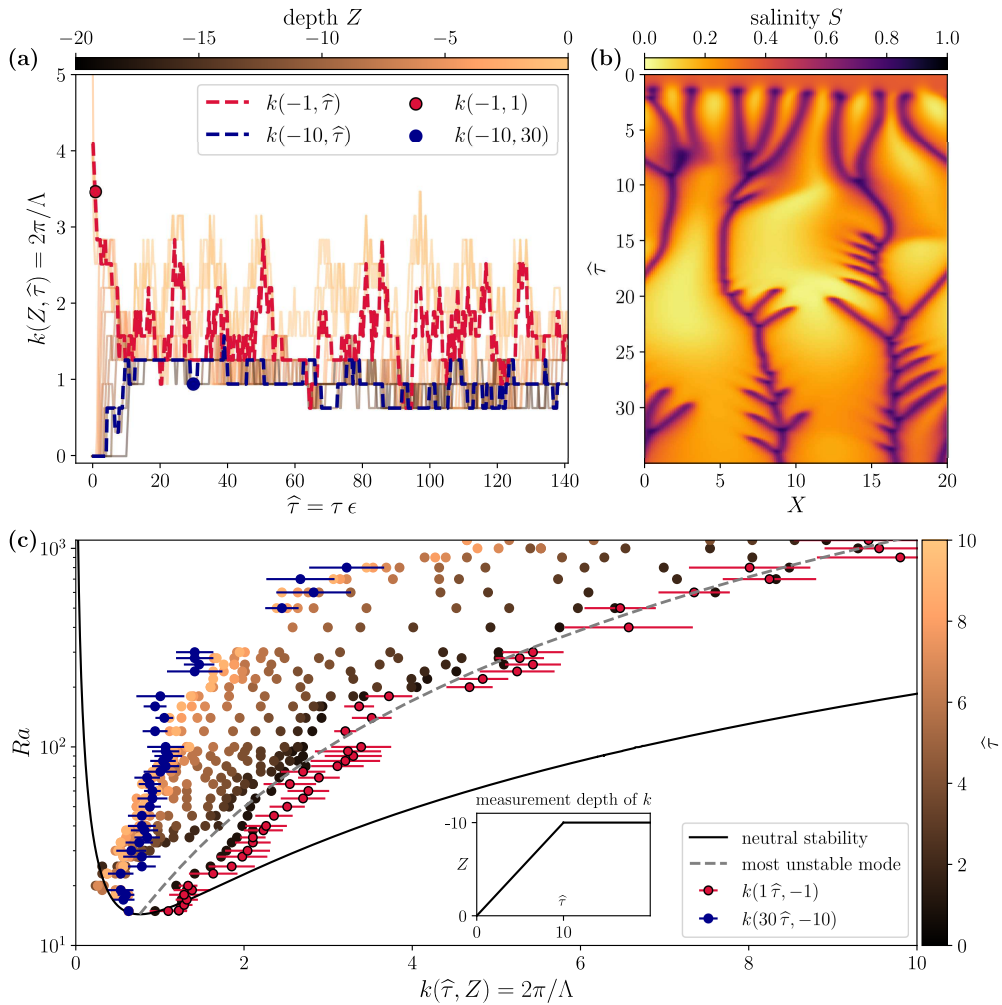}
    \caption{Time-evolution of the plume wavelength, $\Lambda$, and wavenumber, $k$.  The observed wavenumber depends on the depth $Z$ and time $\widehat\tau$ where it is measured, as shown in  \textbf{(a)} for a simulation at $\Ray=100$.  Here, the red dashed curve gives $k$ as measured close to the top boundary, at $Z=-1$.  This metric captures the initially high $k$ values before coarsening occurs, but also fluctuates at later times due to the reinitiation and merging of proto-plumes.  Deeper, the blue dashed curve shows the wavenumber at $Z = -10$, where it has a much more stable value (at least, after the plumes have first descended to this level).  \textbf{(b)} A space-time diagram is shown for the same simulation, where each horizontal line of data represents the state of the simulation for a moment at a depth of $Z=-1$. \textbf{(c)} Over time, all simulations show  coarsening behaviour.  Initially, the $k$ measured near the surface, at $Z=-1$, is similar to the most unstable mode of the linear stability analysis (dashed line). However, as the system ages $k$ decreases and ultimately shows little variation with $\Ray$ in the steady state regime, depicted here for the late-time case of $\widehat\tau = 30$ and at a depth of $Z=-10$.  Data for times and depths intermediate between these two limiting cases are shown as variously coloured circles.   Note that wavenumbers for high Ra and $\widehat{\tau}$ are not available, since simulations at high $\Ray$ were usually terminated before the required duration was reached.}
    \label{fig:wavenumber_development}
\end{figure}

In figure~\ref{fig:wavenumber_development} \textbf{(a)} we show the time development of the plume wavenumber $k$, as measured at different depths for a simulation run at $\Ray= 100$.   The various regimes of the dynamics, described earlier, are reflected in the development of this wavenumber.  The growth, merging and re-initiation of plumes are also apparent in the corresponding space-time diagram displayed in figure~\ref{fig:wavenumber_development} \textbf{(b)}, measured at a depth of $Z=-1$.  At first, both these panels show the linear and flux-growth regimes, with closely spaced plumes of large $k$ and small wavelength.  Choosing a shallow depth of $Z=-1$ and early time of $\widehat\tau = 1$ to characterise the initial response, we find that the wavenumber of $k=3.55\pm0.54$ seen at that time is similar to the most unstable mode of the linear stability analysis, $k=3.01$.  

After the first plumes have more fully formed, at intermediate times the wavenumber of the simulations declines as these plumes begin to strip the boundary layer of salinity and then merge into larger structures.  For the $\Ray=100$ simulation shown in figure~\ref{fig:wavenumber_development} \textbf{(a,b)} this happens between approximately $\widehat{\tau} = 1$ and 10.  From $\widehat{\tau}\approx 10$ onward the plume wavenumber measured deeper into the simulated domain approaches a stable value of $k \approx 1$ (see blue dashed curve in figure~\ref{fig:wavenumber_development} \textbf{(a)} for the example of $Z = -10$). The wavenumber closer to the top boundary (\textit{e.g.} red dashed line in figure~\ref{fig:wavenumber_development}~\textbf{(a)}, for $Z = -1$), fluctuates between that value and a higher value of $k \approx 3$.  This is indicative of the episodic re-initiation of proto-plumes that can also be seen by the herring-bone pattern in the space-time diagram of  figure~\ref{fig:wavenumber_development}~\textbf{(b)}.  These proto-plumes are usually ephemeral, and merge into a larger plume before they can reach into, and be noticed at, the larger depths.  A similar process of intermittent plume initiation was seen in the related model (\textit{i.e.} without evaporation) studied by \cite{SLIM}, who referred to them as proto-plume pulses. 

In order to quantify the coarsening of the plume wavelength more generally, we measured $\Lambda$ and $k$ for a range of conditions, with results given in figure~\ref{fig:wavenumber_development} \textbf{(c)} for simulations with Rayleigh numbers between $14.9$ and $1900$.  For a limiting case that is near the onset of the initial instability, we looked at the early ($\widehat\tau=1$) and near-surface ($Z=-1$) response in all simulations.  As shown in  figure~\ref{fig:wavenumber_development} \textbf{(c)} by the red filled circles, these values closely follow the theoretical prediction for the most unstable mode (grey line, from figure~\ref{fig:neutral_stability_and_growth_rates} \textbf{(a)}), although they tend to slightly exceed it.  As the system ages, however, the dependence of the plume spacing on $\Ray$ weakens.  We tracked this behaviour by measuring $k$ at progressively later times and lower depths (where values are less volatile), as shown by the data in figure~\ref{fig:wavenumber_development} \textbf{(c)}, with the various measurement depths indicated by the figure inset. These results demonstrate how, for a wide range of $\Ray$, the long-time limit of the plume wavelength appears to gradually approach the value that would occur at the critical point, namely $k_\mathrm{c} \simeq 0.76$.  This is highlighted by the blue circles in figure~\ref{fig:wavenumber_development} \textbf{(c)}, which show the wavenumbers as measured at $\widehat\tau = 30$ and a depth of $Z=-10$. We will argue in the following section that the relative independence of $k$ on $\Ray$ results from the depletion of the salt-rich upper boundary layer by plume formation, which effectively reduces it to the thickness of a system forced just beyond its critical point.  

\subsection{Dynamics of the high-salinity boundary layer}

The emergence of plumes in our model is driven by the negative buoyancy of the salt in the diffusive boundary layer near the surface of the dry salt lake. Therefore a closer look at the dynamics of the effective thickness of this layer is warranted, along with its link to how the mature plume wavelength emerges.  

\begin{figure}
    \centering
    \includegraphics[width=\textwidth]{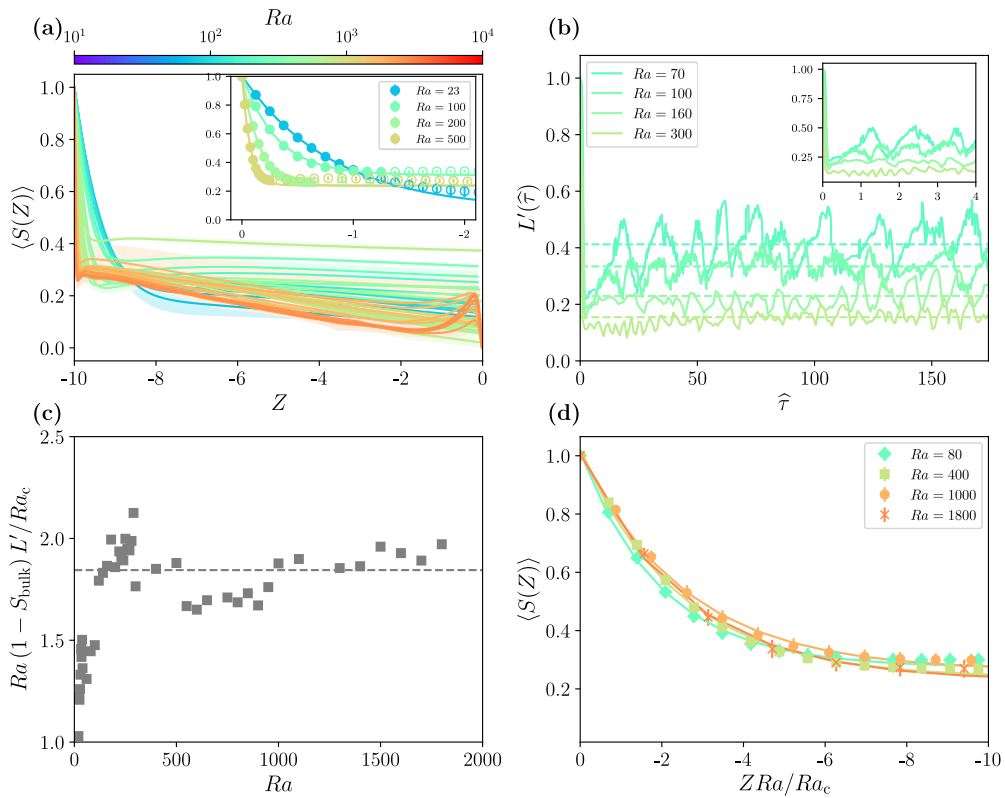}
    \caption{The dynamics of the salinity distribution and its effective boundary layer thickness, $L'$, depend on $\Ray$. \textbf{(a)} The horizontally-averaged salinity, $\left<S(Z)\right>$, is shown for a range of $\Ray$. Results are averaged over 5-10 ensembles for every $\Ray$ and taken from times within the dynamic steady state. The inset shows how an exponential fit is made to the near-surface results (filled dots), to determine $L^\prime$. \textbf{(b)} This thickness evolves over time and in the steady state regime the boundary layer fluctuates around an average value (dashed lines) that depends on $\Ray$. Here, each data set corresponds to a single simulation run.  The inset highlights the initial evolution of the boundary layer during the linear growth and flux growth regimes. \textbf{(c)} The value of $L^\prime$ in the dynamic steady state is inversely proportional to $\Ray$, over a wide range of conditions. In other words, in this regime the buoyancy forces available in the boundary layer are always only slightly above what is required to trigger an instability.  \textbf{(d)} Data collapse showing the salinity distribution in the boundary layers of simulations in their steady state, scaled by the Rayleigh number. }
    \label{fig:length_scaling}
\end{figure}

We have already shown several instances of where plumes drain the boundary layer of solute, thereby limiting the convective drive: compare the salinity distributions near the upper boundaries in figures~\ref{fig:simulation_snapshots} \textbf{(a)} and  \textbf{(h)}, for example. This trend can also be seen in figure~\ref{fig:length_scaling} \textbf{(a)}, where we show the horizontally averaged salinity distributions, $\left<S(Z)\right>$, as measured in the dynamic steady state regime of simulations at different $\Ray$.  Near the surface these results all demonstrate a rapid decay of the salinity with depth, which is stronger for higher $\Ray$. Figure~\ref{fig:length_scaling} \textbf{(a)} also shows how, just below this boundary layer, and especially for higher $\Ray$, the salinity may also pass through a small local maximum of around $S\approx 0.3$.  A similar salinity peak can be noticed in \cite{SLIM} (figure 3). At intermediate depths, below $Z \approx -2$, the salinity then either approaches a constant value or gradually trails off. We note, however, that some of the apparent difference in internal structure seen at these lower depths may simply be due to the restricted height of the simulations at higher $\Ray$ (see appendix A, table~\ref{tab:simulation_stats}).  Similarly, for very high $\Ray$ the salinity starts building up below $Z\approx -9$, since the bottom boundary of the simulated domain is only at $Z=-10$ in these cases.

Given the shape of the horizontally-averaged salinity distribution seen in both the initial and mature states of our simulations, we estimated the \textit{effective} boundary layer thickness at various times and $\Ray$.  To this end, we fit an exponential decay function $S(Z) = \Delta S\, \exp(Z/L^\prime) + S_{\mathrm{bulk}}$ to the rapidly decaying part of the salinity distribution found just below the top boundary, as demonstrated in the inset to figure~\ref{fig:length_scaling} \textbf{(a)}.  For reference, in the stationary solution to Eqs.~\eqref{eq:incompressibilitynondim} to \eqref{eq:darcynondim} the relative salinity \mbox{$S = \exp(Z)$}, so for our initial conditions $L^\prime=1$.  The results for the evolution of $L^\prime$ with time are shown in figure~\ref{fig:length_scaling} \textbf{(b)}.

As with the other metrics discussed here, the width of the salinity boundary layer evolves in different ways as the simulation passes through its various regimes.  Initially, in the linear growth phase the perturbations are not large enough to affect the salinity field, and $L^\prime\simeq 1$ for short times (inset to figure~\ref{fig:length_scaling} \textbf{(b)}).  For higher $\Ray$ the growth rate of these perturbations is faster, so this initial regime is shorter (again, this speeding up is well-captured by the use of $\widehat\tau$ for time).  As the instability enters the flux growth regime, the boundary layer then shrinks as solute is removed by the growing plumes.  Eventually this process stabilises and, as the plumes merge into fewer but larger structures, reverses itself while the boundary layer between the plumes is recharged (we note that this transition can also be related to the velocity overshoot shown in figure~\ref{fig:velocity_scaling} \textbf{(b)}).  Then, when $L^\prime$ becomes sufficiently large it allows for the reinitiation of proto-plumes, which repeat the cycle of growing in amplitude, depleting the boundary layer and merging into larger plumes.  Thus, in the dynamic steady state of the simulations the boundary layer thickness fluctuates around a value of $L'< 1$, corresponding to the episodic emergence of proto-plumes.  The frequency of these proto-plume pulses, and the average value of $L^\prime$, depend on the $\Ray$ of the system: for higher $\Ray$ the boundary layer is thinner and the fluctuations are more rapid.   

For convection driven by one side, \cite{SLIM2013} suggested that (in our notation) an effective Rayleigh number of $\Ray (1-S_{\mathrm{bulk}} )L^\prime$ would describe a boundary layer that had been disturbed by, for example, loss of material to downwelling plumes.  Essentially, $(1-S_{\mathrm{bulk}} )L^\prime$ gives the relative buoyancy forces available in the boundary layer, whereas $\Ray$ characterises the system's general ability to act on those forces.   In all our simulations, the thickness of the effective boundary layer, $L^\prime$, appears to scale approximately with $1/\Ray$ in the steady state regime.   More specifically, as shown in figure~\ref{fig:length_scaling} \textbf{(c)}, the ratio $\Ray (1-S_{\mathrm{bulk}} )L^\prime/\Ray_\mathrm{c}$ approaches a constant value of about 1.8 for $\Ray$ above about 100.   As further evidence of this scaling, figure~\ref{fig:length_scaling}\textbf{(d)} shows a data collapse of the near-surface salinity distributions, $\left<S(Z)\right>$, as rescaled by $\Ray/\Ray_\mathrm{c}$.   These results all suggest that the proto-plume pulses continuously trim back the boundary layer and maintain it in a state that is close to, but just above, a critical condition.  This is consistent with the tendency of the mature plume wavenumber to stabilise at a value near what is expected at these conditions, namely $k_c$.  

\subsection{Effects of varying the initial conditions}\label{subsec:efects_of_varying_the_initial_conditions}

\begin{figure}
    \centering
    \includegraphics[width=\textwidth]{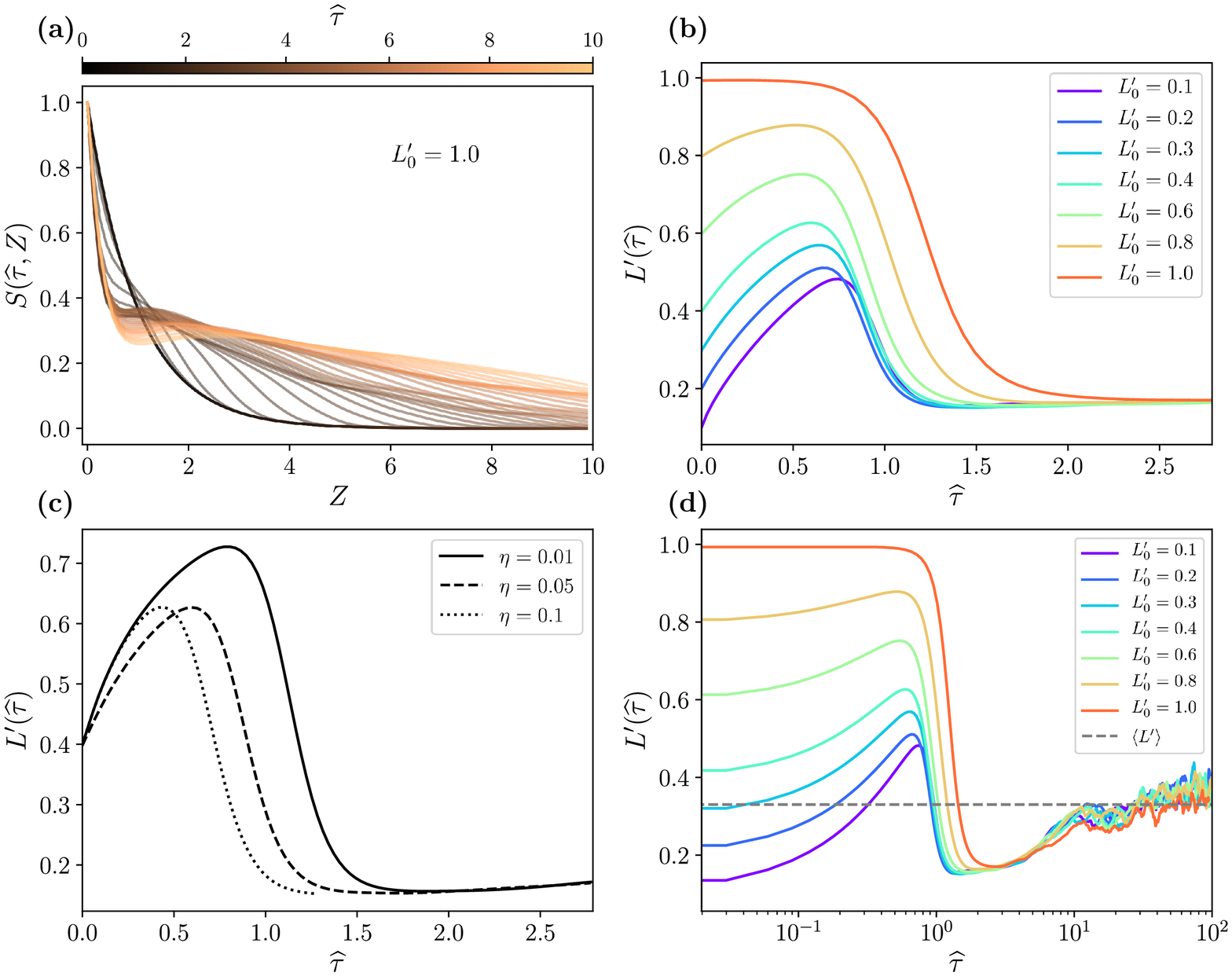}
    \caption{Development of the salinity boundary layer for different initial conditions.  In all cases, salinity distributions were averaged horizontally and over ensembles of 9-11 simulations run at $\Ray = 100$. \textbf{(a)} Development of the near-surface salinity distribution in a simulation with our typical initial condition of $L^\prime_0 = 1.0$, showing its approach to a statistically steady state. \textbf{(b)} Early-time development of the boundary layer thickness $L'$ for different initial conditions $L^\prime_0$. \textbf{(c)} Development of $L^\prime$ starting with $L^\prime_0=0.4$ for different initial noise amplitudes $\eta$. \textbf{(d)} Development of $L^\prime$ for simulations started at the same initial conditions as in \textbf{(b)}, but showing longer timescales. Values converge towards the corresponding steady state value (dashed line) seen in figure~\ref{fig:length_scaling} \textbf{(b)}.}
    \label{fig:initial_conditions}
\end{figure}

Finally, we look at how different initial conditions will modify the system's approach to a mature convection pattern and demonstrate that the dynamic steady state discussed above is robust.  Up to this point our simulations have begun with an initial condition that can be defined as $S = \exp(Z/L_0^\prime)$, where $L_0^\prime=1$, and with random perturbations characterised by an amplitude of $\eta = 0.05$ (see appendix A, Eq.~\ref{eq:randomnoise}).   Now, we varied both the depth of the boundary layer used as the initial salinity distribution and the perturbation amplitude.  For this parameter study we used a somewhat restricted domain, extending down to only $Z=-20$.  The exemplary case of $L_0^\prime=1$ is shown in figure~\ref{fig:initial_conditions} \textbf{(a)}, which has similar features to those discussed at length in regards to figure~\ref{fig:length_scaling}.  We note that at late times here there is an additional slight upward drift of the salinity throughout the domain, which can be explained by a gradual saturation of the salinity in the system once the plumes start interacting with the lower boundary of the simulation.  

In figure~\ref{fig:initial_conditions} \textbf{(b)} we show the development of the effective boundary layer $L^\prime$ for a selection of initial conditions $L_0^\prime$.   Similarly, in figure~\ref{fig:initial_conditions} \textbf{(c)} we show the effect of varying the initial level of noise in our simulation, for a case where $L_0^\prime=0.4$.  When $L^\prime_0<1$ the boundary layer first grows towards the stationary solution of $L^\prime=1$ and then shrinks again, as the convective instability sets in.  Additionally, for lower levels of noise the instability takes longer to manifest itself, allowing the boundary layer slightly more time to be established.   As such, the maximum value of $L'$ depends on the initial conditions, in line with the competition between the growth of the boundary layer and the growth rate of the most unstable mode. These processes are, respectively, a roughly exponential relaxation of the boundary layer thickness (consistent with the fact that the stationary solution is stable to long-wavelength perturbations, see figure~\ref{fig:neutral_stability_and_growth_rates}) and an exponential growth of the first plumes.  Thus, the crossover time between these two processes is relatively insensitive to the initial conditions, and occurs around $\widehat\tau\approx 1$. 

For all these cases, the simulations converge towards the same salinity distributions above times of $\widehat\tau\approx 2$.   This is emphasised in figure~\ref{fig:initial_conditions} \textbf{(d)}, which gives the late-time behaviour of $L^\prime$ for the simulations with different initial boundary layer thicknesses.  Here, all simulations show the same value of $L^\prime$ in the dynamic steady state (although, the smaller domain size adds a slight saturation of $S$ as compared to figure~\ref{fig:length_scaling}).   As described earlier, in this regime the boundary layer thickness self-organises into a state that balances the initiation and growth of new plumes with their coarsening and merging into larger structures.   

\section{Effects of spatially varying evaporation rates}
\label{sect:nonuniform}

Our model of convection in dry salt lakes is inspired by the polygonal patterns that are often seen in the salt crusts at their surface and we will end this work with a discussion of the possible interaction of this crust with the convective plumes.  We first recall that the model also predicts the salinity flux into such a surface crust, which follows from Eq.~\eqref{dim2}.  In dimensional terms, it can be given by $E-\varphi D\partial_z S$.  This flux vanishes in the stationary solution corresponding to our initial condition, $S = \exp(ZE/\varphi D)$, although this scenario will still leave a constant upward flux of \textit{salt} into the crust at whatever concentration is supplied by the reservoir (since $S=0$ corresponds to the background density $\rho_0$, see Eq.~\eqref{eq:salinity}).  By similar argument, when there are convective plumes there should  be less salt flux into the crust above an upwelling--where $\partial_zS$ is high--than above a downwelling.  

A fully dynamical crust, with a thickness varying in response to the salinity flux from the convection beneath it, is challenging to model.  In particular, realistically simulating the evaporation rate can be difficult (see section~\ref{nature} below), but an aspect of this problem that we intend to address in the future.   In this section we will briefly consider, instead, the other side of how a feedback between crust patterns and convection patterns could work.  In other words, we will look at whether periodic variations in the properties of a salt crust could influence any convection pattern occurring beneath it. Since evaporation is the ultimate driver of the dynamics, this allows us to consider the effect of a salt ridge through its control over the local evaporation rate at its location.  In particular, we ask whether particular wavelengths of surface features could stabilise plume locations, leading to potential for long-term feedback between subsurface flows and the crust pattern.     

\begin{figure}
    \centering
    \includegraphics[width=\textwidth]{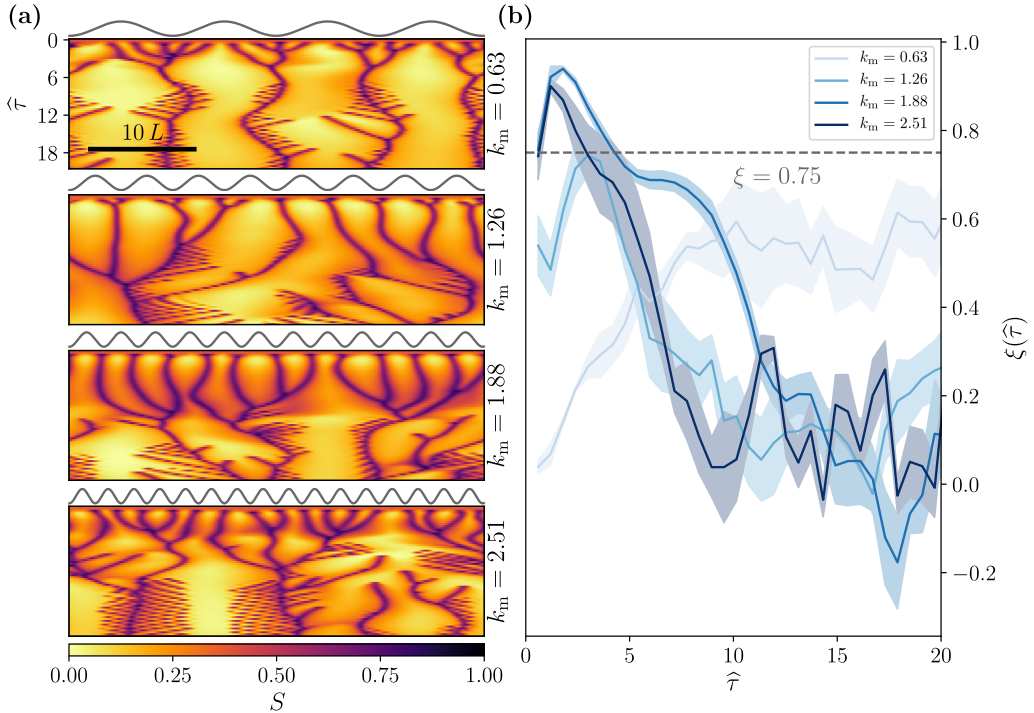}
    \caption{Effect of a modulated evaporation rate on convection.  \textbf{(a)} Space-time diagrams are shown for four simulations at $\Ray=100$ but with surface evaporation rates modulated at an amplitude of $A_\mathrm{m} = 1$ and (from top to bottom) wavenumbers $k_\mathrm{m} = 0.63$, 1.26, 1.88 and 2.51. These give the time-evolution of the salinity profiles at a depth of $Z=-1$.  In each case the evaporation rate at $Z=0$ is indicated by the sine wave on top of the diagram; the high-salinity downwellings show a preference for areas of lower evaporation.   \textbf{(b)} Development of the order parameter $\xi$ with re-scaled time $\widehat{\tau}$ for the same four $k_\mathrm{m}$, also at a depth of $Z=-1$, and averaged over ensembles of five simulation runs.}
    \label{fig:pinning}
\end{figure}

To this end, we modified our model by modulating the surface evaporation rate, such that $U_Z = 1-A_\mathrm{m}\cos{(k_\mathrm{m}X)}$ at $Z=0$, and where $A_\mathrm{m}$ and $k_\mathrm{m}$ are the amplitude and wavenumber of this modulation, respectively (see appendix A for details of how this is implemented).   As the average evaporation rate remains unchanged by the modulation, this change in boundary conditions does not affect the system-averaged Rayleigh number, merely the local conditions.  Space-time diagrams for simulations where $\Ray=100$, $A_\text{m} = 1$ and for four different modulation wavenumbers $k_\mathrm{m}$ are shown in figure~\ref{fig:pinning} \textbf{(a)} and movies of the corresponding simulations are given in supplementary movies S4 through S7.  Several things can be noticed in these simulations.  First, in all cases there is a preference for downwelling plumes to originate at the minima of the evaporation profile.  Second, in most cases the plumes then remain locked at the positions of these minima for some time, before the order breaks down and the dynamics return to a state which resembles a system with a uniform boundary condition.   Third, however, when the wavelength of the evaporation modulation is close to $k_\mathrm{c}\approx0.76$ there is a tendency for the larger plumes to remain trapped near the spots of lower evaporation, even in the dynamic steady state.  

To quantify these points further, we introduce an order parameter that characterises how co-aligned the plumes and evaporation patterns are.  For $N$ plumes this is given by 
\begin{align}
    \xi = - \frac{1}{N}\; \sum_{i=1}^N \cos\left(k_\mathrm{m} X_i\right),
\end{align}
where $X_i$ is the horizontal position of plume $i$, as determined by a minimum in the salinity at a depth of $Z=-1$.   If $\xi = 1$ then the plumes are perfectly aligned with minima in the surface evaporation, whereas when $\xi=-1$ the plumes all lie below maxima and if there is no preferred arrangement of plumes, then $\xi = 0$.  A large $\xi$ would then suggest a route for selecting preferred wavelengths in the crust features, with downwelling plumes trapped by lower-evaporation ridges, and sustaining them with enhannced salinity flux. 

In figure~\ref{fig:pinning} \textbf{(b)}, we show the development of $\xi$ with time for simulations with the same parameters as in figure~\ref{fig:pinning} \textbf{(a)}.  This plot confirms that, especially for larger $k_\mathrm{m}$, the initial plumes start very well-aligned with the surface modulation, with $\xi$ up to about 0.9 at early times.  Since $\Ray = \mathcal{V}_\mathrm{B}/E$, lower evaporation regions will appear, locally, as having a higher \textit{effective} $\Ray$, and it makes sense to expect a higher growth rate of plumes there. The figure also shows that the duration of the initial pinning of plumes under the evaporative minima depends on the modulation wavenumber $k_\mathrm{m}$; \textit{i.e.} the surface modulation can delay the transition from the flux-growth to the merging regime.  Furthermore, in all cases the long-term limit continues to show at least weak ordering, with $\xi$ fluctuating around positive values of about 0.1-0.2 or greater. The $k_\mathrm{m}=0.63$ case shows a marked contrast, however.   Here, the surface modulation is well-matched to the wavelength seen in the dynamic steady state of the simulations and the rise of $\xi$ to about $0.6$ documents how initially disordered plumes arrange themselves to synchronise with the spacing and phase of the modulation. 

\begin{figure}
    \centering
    \includegraphics[width=\textwidth]{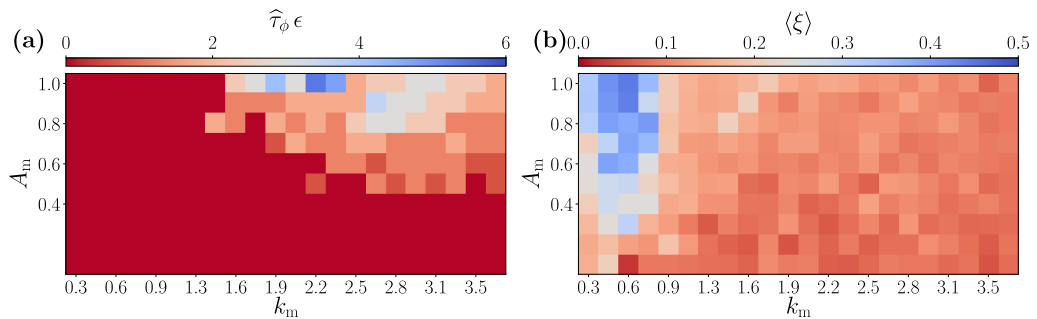}
    \caption{Effects of surface evaporation modulation at \textbf{(a)} early and \textbf{(b)} late times, for simulations at $\Ray=100$ and averaged over ensembles of 5 realisations for each set of conditions.  Plume locations are determined by maxima in the salinity at a depth of $Z=-1$.    \textbf{(a)} The duration of any initial phase-locking, $\widehat\tau_\phi$, depends on the modulation amplitude, $A_\mathrm{m}$, and wavenumber, $k_\mathrm{m}$. In particular, high-amplitude modulation at wavenumbers between 1.9-3.1 can noticeably delay plume merging. \textbf{(b)} Long-term behaviour of the average order parameter $\left<\xi\right>$ for different modulation amplitudes and wavenumbers. This was calculated as the time-averaged value of $\xi$ in the range of $30 \leq \widehat{\tau} \leq 60$. Surface evaporation modulation with $k_\mathrm{m}$ between 0.3 and 0.9 is particularly effective at pinning downwelling plume locations, even down to relatively modest amplitudes of $A_\mathrm{m}=0.3$.}
    \label{fig:synchronization_colormap}
\end{figure}

Figure~\ref{fig:synchronization_colormap} shows how the above results hold true for a wide range of $k_\mathrm{m}$ and $A_\mathrm{m}$. Here, the choice of modulation wavenumbers is limited such that the system width is a multiple of the modulation wavelength. Within this constraint, we chose to investigate a range of modulation wavenumbers in an area of the parameter space that promised to exhibit interesting behaviour, including the full range of wavenumbers seen in the dynamical simulations of $\Ray=100$ for homogeneous boundary conditions (see figure~\ref{fig:wavenumber_development} \textbf{(c)}). For the early-time ordering we define a phase-locking time, $\widehat\tau_\phi$, as the time it takes for the order parameter to drop below $\xi = 0.75$.   Note that this is an arbitrary value, but results are similar for other parameter choices of how to characterise the early-time synchronisation of plumes with flux patterns. Figure~\ref{fig:synchronization_colormap} \textbf{(a)} shows how this time depends on the modulation details.  The initial phase-locking is strongest, i.e. the plumes stay aligned with the minima in the surface modulation the longest,  for modulation wavenumbers between $k_m=1.6$ and $k_m=3.1$.  (\textit{n.b.} a rapid period-doubling instability can be seen in figure~\ref{fig:pinning} \textbf{(a)} for $k_\mathrm{m}=1.26$, suggesting how this alignment breaks down at low $k_\mathrm{m}$). As might be expected, the modulation wavenumbers for which the initial plumes stay aligned the longest roughly corresponds to the wavenumbers first seen to be unstable in the simulations with homogeneous evaporation rates: for $\Ray=100$ this is $k\approx 3.1$. 

To measure the long-time synchronisation of the plumes with the surface modulation we instead averaged the order parameter $\xi$ from $\widehat\tau=30$ to 60. Figure~\ref{fig:synchronization_colormap} \textbf{(b)} shows how this average order parameter, $\left<\xi\right>$, depends on the modulation details. Similar to the phase-locking time, the strength of the ordering increases with the modulation amplitude $A_\mathrm{m}$. More interestingly, $\xi$ exhibits a pronounced maximum in the wavenumber range of $k_\mathrm{m}=0.3$ to $0.9$, which is already present for moderate amplitudes of $A_\mathrm{m}=0.3$. This range broadly matches the critical wavenumber $k_\mathrm{c} \approx 0.76$ as well as the wavenumbers of polygonal salt crust patterns observed in nature, where $k_\mathrm{crust} = 0.78\pm0.43$ with no observable dependence on $\Ray$~\citep{LASSER2019, lasser2020surface}. 

\subsection{Evaporation rate modulation in nature}\label{nature}
In the following we will briefly show how a modulation of the evaporation rate could be realised in the setting of a salt desert with salt ridges in a surface crust.  Although the effect of a salt crust on evaporation can be hard to evaluate exactly (see \textit{e.g.} \cite{Eloukabi2013,Nield2016,Bergstad2017,Farhat2018,Nachshon2018}), the temperature and relative humidity are important parameters, next to salt concentration and air movement, controlling evaporation from a salt pan.  We attempted to estimate the influence of ridges on the micro-climate at the crust level by embedding sensors within salt polygons found in Owens Lake (CA) and tracking the temperature and relative humidity between November 26$^{th}$ and December 2$^{nd}$, 2016.  We used \texttt{HiTemp140} and \texttt{RHTemp1000IS} data loggers, which recorded temperatures and relative humidity every two minutes with a precision of $\pm0.01\,^\circ$C and $\pm0.1\,$\%, respectively. The resulting data are deposited on a public repository~\citep{TEMPHUM} and the protocol for their collection is described in more detail by~\cite{lasser2020surface}, along with further details of the field site.

\begin{figure}
    \centering
    \includegraphics[width=\textwidth]{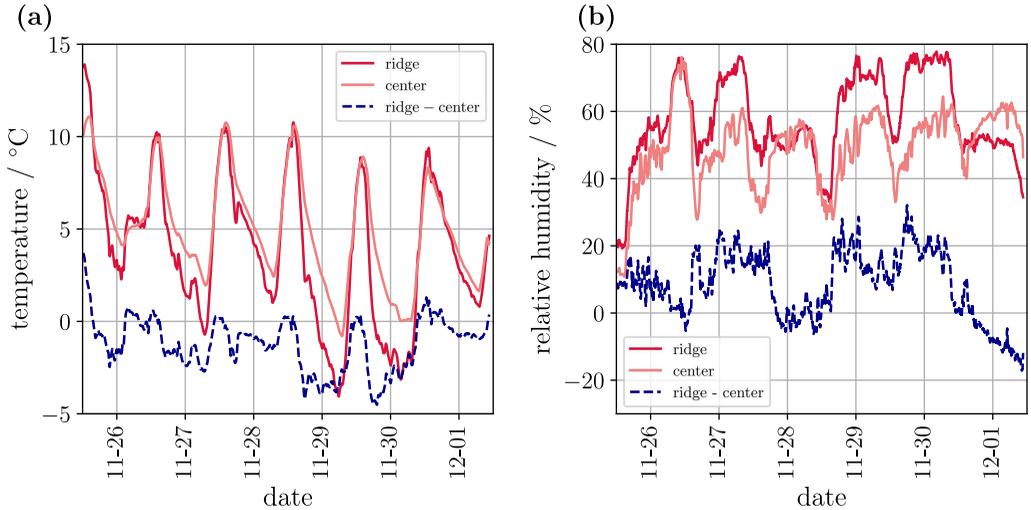}
    \caption{Temperature \textbf{(a)} and relative humidity \textbf{(b)} measurements from salt polygons at Owens Lake, California, conducted in winter 2016. Sensors were embedded either within a ridge or in the centre of a nearby polygon. Displayed values are averaged over 20 minute intervals.}
    \label{fig:field_temp_hum}
\end{figure}

In figure~\ref{fig:field_temp_hum} we show temperature and relative humidity measurements from sensors placed inside a salt ridge and within the crust at the centre of a polygon. In figure~\ref{fig:field_temp_hum} \textbf{(a)} the diurnal fluctuations of temperature between about $10\,^\circ$C at 2:00 PM and $0\,^\circ$C at 6:00 AM are clearly visible. We note that these temperature changes are unlikely to directly affect the density-driven flows by thermal expansion, as they would induce a density change of no more than 1 kg/m$^3$. For comparison, we measured the water immediately below the crust to have a density of at least 200 kg/m$^3$ higher than that at depths of about 1 m \citep{lasser2020surface}.  This difference in magnitude is what justified our original assumption (see Section 2) of ignoring thermal contributions to fluid density, and double-diffusive effects.  These periodic daily fluctuations are also fast compared to the growth of the crust, which occurs over weeks-to-months \citep{NIELD2015}.  Additionally, the highest temperatures recorded during the day are similar between the centre of the crust polygon and the ridge. 

Temperatures inside the ridge, however, drop faster and to lower values at night; the temperature difference is about $2\,^\circ$C on average. 
Figure~\ref{fig:field_temp_hum} \textbf{(b)} shows the development of the relative humidity over the same period. The relative humidity measured inside the ridge is about equal (on the first, third and sixth day) or up to $15\,$\% higher (on the second, fourth and fifth day) than in the polygon centre. Humidity differences are most pronounced during nights and mornings but are preserved to some extent over the course of the day. The difference in relative humidity can be explained by the trapping of moist air below the ridges (see \cite{Nachshon2018} for a discussion of how trapped, stagnant air can also reduce evaporation through salt crusts). Both reduced temperature and increased relative humidity inhibit evaporation from the surface area below a salt ridge and could serve as part of the feedback mechanism proposed above, which modulates the evaporation rate.

In a similar geographic setting a temperature reduction of $1^\circ$C and relative humidity increase of approximately $20\,$\% resulted in a decrease of the evaporation rate from $3\,$mm/day to $2\,$mm/day, or about $30\,$\%~\citep{Farhat2018}. These differences are comparable to the temperature and relative humidity differences we measured between a salt ridge and polygon centre. In figure~\ref{fig:synchronization_colormap} \textbf{(b)} we have illustrated that modulation amplitudes of $A_\mathrm{m} \geq 0.3$ are sufficient to cause a significant spatial ordering of the downwelling plumes. Our observations therefore suggest that the modulation of evaporation rates that could reasonably occur in a field setting would be sufficient to influence the plume positions.

\section{Summary and Discussion}

We have presented a linear stability analysis and subsequent numerical study of buoyancy-driven convection in a fluid-saturated porous medium with surface evaporation and where fluid is replenished from a distant reservoir.   This model is inspired by the situation below a dry lake or salt pan and by the possible connection of the convective dynamics below the ground to the emergence of regular polygonal patterns at the surface \citep{LASSER2019,LASSER_DISS2019}. When rescaled by the evaporation rate $E$ and a length scale that balances advection and diffusion, $L=\varphi D/E$, the model is controlled by a single dimensionless group, the Rayleigh number $\Ray$.  In this context, $\Ray$ can be interpreted as the speed at which a large blob of salt-rich fluid would naturally descend, $\mathcal{V}_\mathrm{B}$, relative to the upward flux of fluid required to balance evaporation.  

There is a stationary solution for the resulting system of equations, corresponding to a salt-rich boundary layer of fluid, of thickness $L$, lying just below the evaporating surface.  Our linear stability analysis considers whether this solution is stable or not, and  complements earlier work concerning the onset of convection in equivalent models \citep{WOODING1960a,HOMSY1976,WOODING1997a,VANDUJIN2002}.  In particular, we confirmed the critical conditions and neutral stability curves given in those works for the surface boundary conditions of either constant evaporation or constant fluid pressure.  Additionally, our analysis extends on previous approaches by solving for the growth rate of an arbitrary small-amplitude perturbation at any Rayleigh number.  Through these methods the initial growth of convective plumes near the evaporating surface was shown to generally be a type-I/finite-wavelength instability, where the most unstable mode increases with increasing $\Ray$. 

In order to follow the evolution of the convective instability past its initial stages, we then performed a range of numerical simulations.  The dynamics of the convection in these simulations pass through several regimes: the sequence is similar to one identified by \citet{SLIM} for the related case of porous media convection also driven from one side, but without any evaporation or through-flow.  The plumes initially follow a linear growth regime, where their wavelength closely matches the predictions of the linear stability analysis.  As their amplitude grows, however, they begin to deplete the salt-rich fluid near the surface, and the dynamics pass into a flux-growth regime.  In the subsequent merging regime the system coarsens as the growing plumes begin to interact and join together into larger structures.  This process leaves gaps between plumes and in the re-initiation regime new proto-plumes appear in these gaps, \textit{via} similar instabilities in the boundary layer.  The emerging proto-plumes are quickly drawn into larger and more stable downwelling plumes, allowing for new proto-plume pulses to occur episodically.   We characterised this re-initiation regime as a dynamic steady state of the system.

Throughout this study we focused on exploring how the dynamical properties of the convection scaled with its driving parameters, here summarised by the Rayleigh number or by the related proximity of the system to its critical point, $\epsilon = (\Ray - \Ray_\mathrm{c})/\Ray_\mathrm{c}$.  In the linear growth regime the growth rate of the most unstable mode $\alpha_m\sim \epsilon$, as expected for a type-I instability. Similarly, for later regimes the plume velocities, measured in various ways, were found to be proportional to $\epsilon$.  This follows naturally from the definition of $\Ray=\mathcal{V}_\mathrm{B}/E$, since the velocity scaling of our system is based on the evaporation rate $E$. Given this, we found a timescale $\widehat\tau = \tau \epsilon = \epsilon t E^2/\varphi^2D$ to be a convenient rescaling.  Specifically, the various regime transitions occur at times of order $\widehat\tau=1$ for a wide range of initial conditions and Rayleigh numbers. In terms of length-scales, the most unstable mode of the linear instability increases monotonically with $\Ray$.  However, due to the tendency of plumes to merge, this response is only seen at very short times--we would not expect it to affect the growth of salt crusts in realistic settings.  Instead, in the dynamic steady state we found that the constant interplay between proto-plume initiation and merging resulted in a plume spacing that was largely independent of $\Ray$, such that the wavenumber of the mature plumes always approached the critical wavenumber of $k_\mathrm{c} \approx 0.76$.  

We note that this last result is in contrast to the more well-studied case of \textit{two-sided} convection, where the thickness of the entire convecting domain, $H$, imposes a scaling of the mature wavenumber with a Rayleigh number defined alternatively as $\Ray =\mathcal{V}_\mathrm{B} H/\varphi D$ (\textit{e.g.} \cite{HEWITT2014} who suggest $k\sim\sqrt{\Ray}$ or \cite{FU2014} who suggest $k\sim\Ray$).  In a two-sided system, if a plume naturally moves at a speed $\sim\mathcal{V}_\mathrm{B}$ across the domain height $H$, then a plume spacing of order $1/\sqrt{\Ray}$ reflects the distance over which concentration gradients would diffuse over the course of the plume's fall \citep{Liang2018}.  For our \textit{one-sided} case of convection the system height is irrelevant.  Instead, the boundary conditions provide the length scale, $L$, over which the typical contributions of advection and diffusion balance, and our Rayleigh number can be written as $\Ray=\mathcal{V}_\mathrm{B} L/\varphi D$.  The length $L$ is also the equilibrium thickness of a heavy boundary layer of fluid that would otherwise naturally develop below the surface and so it characterises the potential driving force available for convection.  We argued that the relative independence of the plume spacing with $\Ray$, observed in out simulations, relates to how convection depletes this salt-rich boundary to leave a layer just barely thick enough to allow for convection.  This conclusion was supported by a demonstration that in the steady state regime the effective thickness of the boundary layer, $L^\prime$, is proportional to $1/\Ray$ over a wide range of $\Ray$ and by a data collapse of the shape of the horizontally-averaged salinity field with the same scaling.  Put simply, if the boundary layer grows much thicker than this, then it will favour the rapid growth of new proto-plumes, which will strip the layer back down to close to a critical thickness before they disappear through plume mergers.  We argued that this balance also controls the plume spacing and is why the steady state wavenumber of the convection plumes is always maintained at a value near $k_\mathrm{c}$, regardless of the real Rayleigh number of the system.  

Finally, we modified our model to allow for inhomogeneous boundary conditions, namely a sinusoidal modulation of the evaporation rate in space.  This modulation is a first step for exploring how feedback between subsurface convection and surface crust growth could work.  The influence of ridges on the evaporation rate is supported by data measured in the field, which shows a difference in temperature and relative humidity below ridges, as compared to salt polygon centres.  We found that for a range of modulation wavenumbers and amplitudes, regions of locally suppressed evaporation could pin downwelling plumes in place for long periods of time.  This pinning was particularly apparent for modulations around the critical wavenumber, $k_\mathrm{c}$.   

Put together, our results indicate that the vigorous convection patterns that emerge in our model system are  robust to differences in $\Ray$, for Rayleigh numbers far enough above $\Ray_\mathrm{c}$. Such a robustness to fluctuations in the environment is an important feature of any mechanism driving salt polygon emergence in nature, as these patterns occur in areas with vastly different conditions but nonetheless display remarkably consistent length scales (\textit{e.g.} \cite{CHRISTIANSEN1963,KRINSLEY1970,NIELD2015}).  A plume wavelength of $2\pi/k_\mathrm{c}$ corresponds to features with a spacing of a few meters, assuming diffusion constants of order 10$^{-9}$ m/s$^2$ and evaporation rates of order 1 mm/day (or, 10$^{-8}$ m/s).   This agrees both with the observed sizes of salt polygons and with the scales of convective plumes seen under similar conditions in tidal flats and sabkhas \citep{VANDAM2009,STEVENS2009}.  We intend to develop this argument further with detailed comparisons to field data elsewhere \citep{LASSER2019,lasser2020surface}. Furthermore, although motivated by the problem of a dry salt lake, the model system developed here could also be applied to other cases of porous media convection where there is some background through-flow of fluid across the convecting domain.  \\

\noindent{\bf Acknowledgements\bf{.}}  We thank C\'edric Beaume for discussions and a close reading of the manuscript; Matthew Threadgold for sharing the eigenvalue solver code adapted for Section 3.2; Joanna M. Nield for assistance with field work; and Antoine Fourri\`ere for early discussions on pattern formation mechanisms. \\

\noindent{\bf Declaration of Interests\bf{.}}  The authors report no conflict of interest. \\

\noindent{\bf  Author ORCID\bf{.}}   J. Lasser, https://orcid.org/0000-0002-4274-4580; L. Goehring, https://orcid.org/0000-0002-3858-7295 \\

\noindent{\bf Author contributions\bf{.}} M.E. and L.G. derived the theory, M.E. and J.L. wrote the code and J.L. performed and interpreted the numerical experiments.  J.L. and L.G. performed the field work. All authors contributed to writing the paper. \\

\appendix
\section{Numerical simulation}\label{appA}

We implemented a two-dimensional finite-difference model of an evaporating salt lake, based on the non-dimensional system of equations given by Eqs.~\eqref{eq:incompressibilitynondim}--\eqref{eq:darcynondim}.  This uses a pseudo-spectral approach and a stream function-vorticity representation, similar to \citet{CHEN1998a,RUITH2000,RIAZ2003}, a sixth-order compact finite difference scheme to compute spatial derivatives and an explicit fourth-order Runge-Kutta scheme for time-stepping.  The code has been made available on GitHub \citep{simulationcode}.   

\subsubsection*{Model setup}
The numerical model simulates a two-dimensional $(X,Z)$ area of width $W$ in the $X$-direction and height $H$ in the $Z$-direction. For the salinity $S$ and fluid flux $\vec{U} = (U_X,U_Z)$ we assume periodic boundary conditions in the $X$-direction.  The relative salinity $S = 1$ at the top boundary ($Z=0$) and $S = 0$ at the lower boundary ($Z=-H)$.  At $Z=0$ surface evaporation is modelled as a boundary condition on $U_Z$.  To allow for a modulation in evaporation rate, $U_Z = 1-A_\mathrm{m}\cos{(k_\mathrm{m}X)}$ there, where $A_\mathrm{m}$ represents the strength of the modulation and $k_\mathrm{m}$ its wavenumber; for constant evaporation, $A_\mathrm{m}=0$.  In all cases the average value of $U_Z$ at the surface is 1.  At the bottom boundary, $Z=-H$, fluid recharge is assumed to be uniform, such that $U_Z = 1$ there.

We use potential functions to reformulate the equations of porous media flow, following \citet{RUITH2000, RIAZ2003}. For a two-dimensional and incompressible flow the flux $\vec{U}$ is related to the Lagrange stream function $\psi$ by $\vec{U} = (\partial_Z \psi,-\partial_X\psi)$. The vorticity, $\omega = \partial_XU_Z-\partial_ZU_X$, is then given by the Poisson equation,
\begin{equation}
    \vec{\nabla}^2 \,\psi = -\omega.
    \label{eq:implPoisson}
\end{equation}
Taking the curl of Darcy's Law, Eq.~\eqref{eq:darcynondim}, allows us to eliminate the pressure term, as $\nabla\times(\nabla P) = 0$, and solve for the vorticity as
\begin{equation}
    \omega = - \Ray \partial_XS \label{eq:DarcyCurl}.
\end{equation}
The salt mass balance of Eq. \eqref{eq:saltconservationnondim} can also be written in terms of the stream function:
\begin{equation}
    \partial_\tau S  = (\partial_X\psi)(\partial_Z S) - (\partial_Z\psi)(\partial_X S) + \partial_X^2 S+\partial_Z^2 S. \label{eq:saltconservationimpl}
\end{equation}

To solve Eqs.~\eqref{eq:implPoisson}--\eqref{eq:saltconservationimpl} we need boundary conditions for the stream function $\psi$ and vorticity $\omega$. The constraints on flow give $\partial_X\psi = -1+A_\mathrm{m}\cos(k_\mathrm{m}X)$ at $Z=0$ and $\partial_X\psi = -1$ at $Z = -H$.  These inhomogeneous boundary conditions can be accounted for by taking $\psi = \psi^\prime + \Psi$, where
\begin{equation}
    \psi^\prime = -X + \frac{A_\mathrm{m}}{k_\mathrm{m}}\sin\left(k_\mathrm{m}X\right)\cos^2\left(\frac{\pi Z}{2 H}\right),
\end{equation}
such that $\partial_X\Psi = 0$ at both $Z=0$ and $Z=-H$.  The constant salinity conditions correspond to a vanishing vorticity, $\omega = 0$, at both these surfaces.  As an initial condition we make use of 
\begin{align}
S_0 = \dfrac{e^{Z} - e^{-H}}{1-e^{-H}},
\label{eq:simsteadystate}
\end{align}
which is the stationary solution of the salinity when $A_\mathrm{m} = 0$.  Following \citet{RIAZ2003} we introduce perturbations by adding random fluctuations into the initial salinity.  To do so we generate random numbers $f(X,Z)$, uniformly distributed in $[-1,1]$, for all grid points.  In order to avoid artefacts in derivatives, these are then convolved with a Gaussian function of width $\sigma = 3$ grid cells, such that 
\begin{align}
  {S}^\prime = \eta \, f(X,Z)\,*\, e^{-\frac{X^2+Z^2}{\sigma^2}},
  \label{eq:randomnoise}
\end{align}
where $\eta$ gives the magnitude of the perturbation.    The initial salinity field is then $S = S_0 + S^\prime$, at $\tau = 0$. Unless otherwise stated we used a default value of $\eta=0.05$. 

\subsubsection*{Implementation}
To numerically solve the governing equations of the simulation we largely follow the implementation described by~\cite{RIAZ2003, RUITH2000}. Specifically, at each time step we: (i) compute derivatives in the $X$-direction by first making use of a Fourier transform; (ii) compute derivatives in the $Z$-direction by using a compact finite difference scheme~\citep{LELE1992}; and (iii) use an explicit fourth-order Runge-Kutta scheme for the time-integration of equation~\eqref{eq:saltconservationimpl}.

The simulation is performed on a grid of $M\times N$ points at positions $(x_m,z_n)$.  The grid spacings $\Delta X$ and $\Delta Z$ are varied with $\Ray$, since the spatial resolution has to be increased at higher $\Ray$ to resolve all relevant features.  Domain sizes were also adjusted, allowing for a similar number of gridpoints in most simulations.  The simulation grid spacings and domain width $W$ and height $H$ are given in table~\ref{tab:simulation_stats} for all $\Ray$.  The time step $\Delta \tau$ follows the CFL-condition \citep{COURANT1928},
\begin{align*}
  C = \frac{U_\textrm{max}\,\Delta \tau}{\Delta X} \leq C_0, 
\end{align*}
where $U_\textrm{max}$ is the maximum speed occurring at that time, and where we set $C_0=0.1$.

\begin{table}
\centering
\begin{tabular}{l|c|c|c|c|c}
$\Ray$ & figure 6 & figure 7 & figure 8 & figure 9 & figures 10 \& 11 \\
\hline
$14.9-20$ & - & $20\times 20$ & - & - & - \\
$20$ & $40\times 100$ & $40\times 100$ & - & - & \\
$23 - 40$ & $20\times 20$ &  $20 \times 20$ & $20\times 20$ & - & - \\
$40$ & $40\times 100$ & $20\times 20$ & $20\times 20$ & - & - \\
$45 - 50$ & - & $20\times 20$ & $20\times 20$ & - & - \\
$60$ & $40\times 100$ & $20\times 20$ & $20\times 20$ & - & - \\
$65 - 75$ & - & $20\times 20$ & $20\times 20$ & - & - \\
$80$ & $40\times 100$ & $20\times 20$ & $20\times 20$ & - & - \\
$85 - 95$ & - & $20\times 20$ & $20\times 20$ & - & - \\
$100$ & $40\times 100$ & $20\times 20$ & $20\times 20$ & $20\times 20$ &$40\times 40$\\
$120 - 140$ & $40\times 100$ & $40\times 100$ & $40\times 100$ & - & - \\
$160$ & $40\times 40$ & $40\times 100$ & $40\times 100$ & - & - \\
$180$ & $40\times 100$ & $40\times 100$ & $40\times 100$ & - & - \\
$200$ & $12\times 25$ & $20\times 40$ & $40\times 80$ & - & - \\
$220 - 290$ & $40\times 80$ & $20\times 40$ & $40\times 80$ & - & - \\
$300$ & $12\times 25$ & $20\times 40$ & $12\times 25$ & - & - \\
$400$ & $10\times 20$ & $13\times 25$ & $10\times 20$ & - & - \\
$450$ & $10\times 20$ & - & $10\times 20$ & - & - \\
$500$ & $10\times 20$ & $10\times 20$ & $10\times 20$ & - & - \\
$550$ & $7\times 15$ & - & $7\times 15$ & - & - \\
$600$ & $7\times 15$ & $10\times 20$ & $7\times 15$ & - & - \\
$650$ & $7\times 15$ & - & $10\times 20$ & - & - \\
$700$ & $7\times 15$ & $8\times 15$ & $7\times 15$ & - & - \\
$750$ & $7\times 15$ & - & $7\times 15$ & - & - \\
$800$ & $5\times 10$ & $8\times 15$ & $7\times 15$ & - & - \\
$900-1000$ & $5\times 10$ & $5\times 10$ & $7\times 15$ & - & - \\
$1100 - 1800 $ & $5\times 10$ & $5\times 10$ & $5\times 10$ & - & - \\
$2000$ & - & - & $5\times 10$ & - \\
\hline
\end{tabular}
\caption{\label{tab:simulation_stats} System dimensions, $W\times H$, of simulations for different $\Ray$ used in Figures~\ref{fig:velocity_scaling},~\ref{fig:wavenumber_development},~\ref{fig:length_scaling},~\ref{fig:initial_conditions},~\ref{fig:pinning} and~\ref{fig:synchronization_colormap}. All domain sizes are given in units of the natural length $L$.}
\end{table}

\emph{Derivatives in \textit{X}-direction}: We employ Fourier expansions, with coefficients
\begin{align}
  \widehat\Psi_{\xi}(z_n) = \dfrac{1}{M}\sum \limits_{m= 1}^{M} \Psi(x_m,z_n) e^{-2\pi i\xi x_m/W} 
   \label{eq:fgexpansion}
\end{align}
and analogous expressions for $\widehat\psi^\prime_{\xi}$ and $\widehat\omega_{\xi}$, where $\xi$ can take integer values between $\pm M/2$.  In terms of these Fourier coefficients Eq.~\eqref{eq:implPoisson} may be written as
\begin{equation}
  \left(\dfrac{2\pi\iu \xi}{W} \right)^2\,\widehat\Psi_{\xi} + \partial^2_{Z}\,\widehat\Psi_{\xi} = - \widehat\omega_{\xi} - \widehat{\omega}_{\xi 0} \label{eq:fourier_intermediate_step},
\end{equation}
where $\widehat\omega_{\xi}$ are calculated at each time step by a similar transformation of Eq.~\eqref{eq:DarcyCurl}, and where $\widehat{\omega}_{\xi 0} = \nabla^2\widehat\psi^\prime_\xi$ are constant in time and arise from the boundary conditions. 

 \emph{Derivatives in $Z$-direction:} To solve the system of equations given by Eq.~\eqref{eq:fourier_intermediate_step} we  computed $\partial_{Z}^2\widehat{\Psi}_\xi (z_n)$ following the implicit sixth-order compact finite difference scheme given by~\citet{CARPENTER1993}. The $N$ linear differential equations can be described by two matrices such that
\begin{align*}
    A_\mathrm{left} \partial^2_{Z} \widehat{\Psi}_\xi(z_n) = A_\mathrm{right} \widehat{\Psi}_\xi(z_n) \quad \Rightarrow \quad \partial^2_{Z} \widehat{\Psi}_\xi(y_n) = \left( A^{-1}_\mathrm{left} A_\mathrm{right}\right) \widehat{\Psi}_\xi(z_n).
    \label{eq:matrix_replacement}
\end{align*}
To construct the matrices $A_\mathrm{left}$ and $A_\mathrm{right}$, we use the coefficients $\alpha_i$, $\beta_i$, $a_i$ and $b_i$ listed in table~\ref{tab:coefficientsTable}, as calculated by \citet{TYLER2007}. At each time step we then solve for the Fourier components $\widehat\psi$, which are inverted to give the stream function $\psi(x_m,z_n)$.  The flux $\vec{U}$ is then calculated from  the stream function by again using an implicit sixth-order compact finite difference scheme for the spatial derivatives. Coefficients for the first and second-order derivatives of interior and boundary points are listed in table~\ref{tab:coefficientsTable}.

\begin{table}
    \centering
    \begin{tabular}{c|cc|cccccc}
        & $\alpha_0$ & $\alpha_1$& $a_1$& $a_2$& $a_3$& $a_4$& $a_5$& $a_6$\\ \hline
        interior & &$1/3$ &  $14/9$ & $1/9$ &&&&\\
        node $0$ & $5$ & & $197/60$ & $-5/12$ & $5$ & $-5/3$ & $5/12$ & $-1/20$ \\
        node $1$ & $3/4$ & $1/8$ & $-43/96$ & $-5/6$ & $9/8$ & $1/6$ & $-1/96$ &\\
    \end{tabular}
    \bigskip \\
    \begin{tabular}{c|cc|cccccc}
        & $\beta_1$ & $\beta_2$ & $b_1$& $b_2$& $b_3$& $b_4$& $b_5$& $b_6$\\ \hline
        interior & $12/97$ & $-1/194$& $120/7$ &&&&&  \\
        boundary & $11/12$ & $-131/4$ & $177/16$& $-507/8$& $783/8$& $-201/4$& $81/16$& $-3/8$ 
    \end{tabular}
    \caption{\label{tab:coefftyler} Coefficients for sixth order compact finite difference schemes used in the numerical simulations, following \citet{LELE1992,TYLER2007}.  The $\alpha$ and $a$ terms are used to calculate first derivatives in $Z$, while $\beta$ and $b$ terms are used for second derivatives.} \label{tab:coefficientsTable}
\end{table}

\emph{Time-integration}: To update the salinity field \textit{via} Eq.~\eqref{eq:saltconservationimpl} we used a fourth-order Runge-Kutta scheme and the coefficients given for example by~\citet[p.~352]{SULI2003}. For spatial derivatives we again use an implicit sixth-order compact finite difference scheme. 

\subsubsection*{Validation}
\begin{figure}
    \centering
    \includegraphics[width=\textwidth]{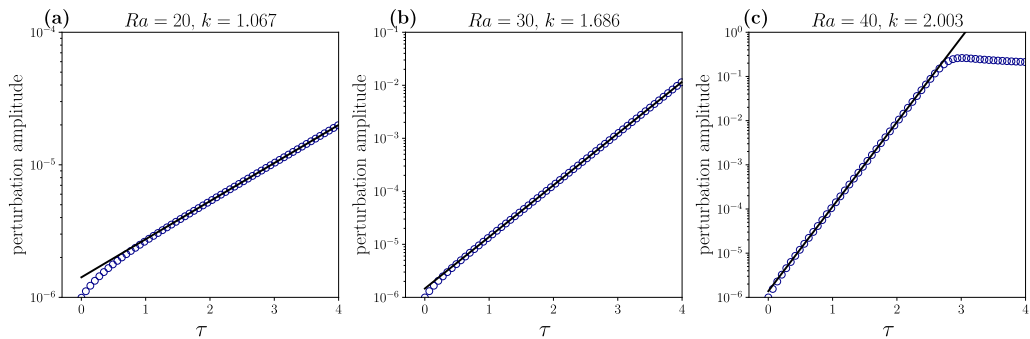}
    \caption{Code was validated for single-mode perturbations of initial amplitude $\eta = 10^{-6}$, on domains of depth $H=12$ and width $W=2\pi/k$.  Growth rates $\alpha$ were obtained by fitting the measured amplitude (blue circles) with an exponential fit (black line).   Examples are given for: \textbf{(a)} $\Ray=20$, $k=1.067$, growth rate of fit $\alpha_{fit}=0.662$, compared to $\alpha=0.660$ from linear stability analysis; \textbf{(b)} $\Ray=30$, $k=1.686$, $\alpha_{fit}=2.245$ compared to $\alpha = 2.250$; \textbf{(c)} $\Ray=40$, $k=2.003$, $\alpha_{fit}=4.407$ compared to $\alpha = 4.414$.}
    \label{fig:amplitude_measurement}
\end{figure}

The model was validated by comparison with the results of the linear stability analysis. A similar approach is used in \citet{RUITH2000, RIAZ2003, CHEN1998a, TAN1988}.  For this, instead of Eq. \eqref{eq:randomnoise} we perturbed our initial conditions with 
\begin{align}
S^\prime = -\eta\,Z\,\left(\dfrac{e^{Z}-e^{-H}}{1-e^{-H}}\right)\cos{\left(kX\right)}, \nonumber
\end{align}
using a perturbation of initial magnitude $\eta = 10^{-6}$.  This perturbation is consistent with the constant salinity boundary condition and affects only a single wavenumber $k$.  Its growth over time was measured from (half) the peak-to-peak amplitude of this mode evaluated at a depth of $Z=-1$. The simulated growth rate $\alpha_{fit}$ was determined by fitting an exponential to the amplitude measurements. For this fit we focus on the linear growth phase, manually excluding any initial transient or later non-linear saturation. Some example measurements, and fits, are shown in figure~\ref{fig:amplitude_measurement}. Results for various $\Ray$ and $k$ agree with the theoretical values to within a relative error of order $1\%$, as was shown in figure~\ref{fig:neutral_stability_and_growth_rates} \textbf{(b)}.  

\bibliographystyle{jfm}
\bibliography{sources}

\begin{thebibliography}{77}
\expandafter\ifx\csname natexlab\endcsname\relax\def\natexlab#1{#1}\fi
\def\au#1{#1} \def\ed#1{#1} \def\yr#1{#1}\def\at#1{#1}\def\jt#1{\textit{#1}}
  \def\bt#1{#1}\def\bvol#1{\textbf{#1}} \def\vol#1{#1} \def\pg#1{#1}
  \def\publ#1{#1}\def\arxiv#1{#1}\def\org#1{#1}\def\st#1{\textit{#1}}

\bibitem[Askey \& Daalhuis(2010)]{askey2010generalized}
{\sc \au{Askey, R.A.} \& \au{Daalhuis, A.B.~Olde}} \yr{2010}  \at{{Generalized
  hypergeometric functions and Meijer G-function}}.  \bt{In {\em NIST Handbook
  of Mathematical Functions\/} (ed. \ed{Frank W.~J. Olver, Daniel~W. Lozier,
  Ronald~F. Boisvert \& Charles~W. Clark})},  \pg{pp. 403--418}.
  \publ{Cambridge University Press}.

\bibitem[Bergstad {\em et~al.\/}(2017)Bergstad, Or, Withers \&
  Shokri]{Bergstad2017}
{\sc \au{Bergstad, Mina}, \au{Or, Dani}, \au{Withers, Philip~J.} \& \au{Shokri,
  Nima}} \yr{2017}  \at{{The influence of NaCl concentration on salt
  precipitation in heterogeneous porous media}}.  \jt{Water Resour. Res.}
  \bvol{53},  \pg{1702--1712}.

\bibitem[Boufadel {\em et~al.\/}(1999)Boufadel, Suidan \& Venosa]{BOUFADEL1999}
{\sc \au{Boufadel, M.C.}, \au{Suidan, M.T.} \& \au{Venosa, A.D.}} \yr{1999}
  \at{Numerical modeling of water flow below dry salt lakes: effect of
  capillarity and viscosity}.  \jt{J. Hydrol.}  \bvol{221}~(1),  \pg{55--74}.

\bibitem[Briere(2000)]{BREIRE2000}
{\sc \au{Briere, Peter~R.}} \yr{2000}  \at{Playa, playa lake, sabkha: proposed
  definitions for old terms}.  \jt{J. Arid Environ.}  \bvol{45}~(1),
  \pg{1--7}.

\bibitem[Brunner {\em et~al.\/}(2004)Brunner, Bauer, Eugster \&
  Kinzelbach]{BRUNNER2004}
{\sc \au{Brunner, Philip}, \au{Bauer, Peter}, \au{Eugster, Martin} \&
  \au{Kinzelbach, Wolfgang}} \yr{2004}  \at{{Using remote sensing to
  regionalize local precipitation recharge rates obtained from the Chloride
  Method}}.  \jt{J. Hydrol.}  \bvol{294}~(4),  \pg{241--250}.

\bibitem[Bryant(2003)]{BRYANT2003}
{\sc \au{Bryant, Robert~G.}} \yr{2003}  \at{{Monitoring hydrological controls
  on dust emissions: preliminary observations from Etosha Pan, Namibia}}.
  \jt{Geogr. J.}  \bvol{169}~(2),  \pg{131--141}.

\bibitem[Bryant \& Rainey(2002)]{BRYANT2002}
{\sc \au{Bryant, R.~G.} \& \au{Rainey, M.~P.}} \yr{2002}  \at{{Investigation of
  flood inundation on playas within the Zone of Chotts, using a time-series of
  {AVHRR}}}.  \jt{Remote Sens. Environ.}  \bvol{82}~(2--3),  \pg{360--375}.

\bibitem[Busse \& Joseph(1972)]{BUSSE1972}
{\sc \au{Busse, F.~H.} \& \au{Joseph, D.~D.}} \yr{1972}  \at{Bounds for heat
  transport in a porous layer}.  \jt{J. Fluid Mech.}  \bvol{54}~(3),
  \pg{521--543}.

\bibitem[Carpenter {\em et~al.\/}(1993)Carpenter, Gottlieb \&
  Abarbanel]{CARPENTER1993}
{\sc \au{Carpenter, Mark~H.}, \au{Gottlieb, David} \& \au{Abarbanel, Saul}}
  \yr{1993}  \at{The stability of numerical boundary treatments for compact
  high-order finite-difference schemes}.  \jt{J. Comput. Phys.}
  \bvol{108}~(2),  \pg{272--295}.

\bibitem[Chen \& Meiburg(1998{\natexlab{{\em a\/}}})]{CHEN1998a}
{\sc \au{Chen, Ching-Yao} \& \au{Meiburg, Eckart}} \yr{1998{\natexlab{{\em
  a\/}}}}  \at{{Miscible porous media displacements in the quarter five-spot
  configuration. Part 1. The homogeneous case}}.  \jt{J. Fluid Mech.}
  \bvol{371},  \pg{233--268}.

\bibitem[Chen \& Meiburg(1998{\natexlab{{\em b\/}}})]{CHEN1998b}
{\sc \au{Chen, Ching-Yao} \& \au{Meiburg, Eckart}} \yr{1998{\natexlab{{\em
  b\/}}}}  \at{{Miscible porous media displacements in the quarter five-spot
  configuration. Part 2. Effect of heterogeneities}}.  \jt{J. Fluid Mech.}
  \bvol{371},  \pg{269--299}.

\bibitem[Christiansen(1963)]{CHRISTIANSEN1963}
{\sc \au{Christiansen, F.W.}} \yr{1963}  \at{{Polygonal fracture and fold
  systems in the salt crust, Great Salt Lake Desert, Utah}}.  \jt{Science}
  \bvol{139}~(3555),  \pg{607--609}.

\bibitem[Courant {\em et~al.\/}(1928)Courant, Friedrichs \& Lewy]{COURANT1928}
{\sc \au{Courant, R.}, \au{Friedrichs, K.} \& \au{Lewy, H.}} \yr{1928}
  \at{{\"Uber die partiellen Differentialgleichungen der mathematischen
  Physik}}.  \jt{Math. Ann.}  \bvol{100},  \pg{32--74}.

\bibitem[Cross \& Greenside(2009)]{CROSS2009}
{\sc \au{Cross, Michael} \& \au{Greenside, Henry}} \yr{2009} {\em Pattern
  Formation and Dynamics in Nonequilibrium Systems\/}.  \publ{Cambridge
  University Press}.

\bibitem[DeMeo {\em et~al.\/}(2003)DeMeo, Laczniak, Boyd, Smith \&
  Nylund]{DEMEO2003}
{\sc \au{DeMeo, Guy~A.}, \au{Laczniak, Randell~J.}, \au{Boyd, Robert~A.},
  \au{Smith, J.~LaRue} \& \au{Nylund, Walter~E.}} \yr{2003} {Estimated
  ground-water discharge by evapotranspiration from Death Valley, California,
  1997--2001}. {U.S. Geological Survey, Water-Resources Investigations Report
  03-4254}.

\bibitem[van Duijn {\em et~al.\/}(2002)van Duijn, Pieters, Wooding \& van~der
  Ploeg]{VANDUJIN2002}
{\sc \au{van Duijn, C.J.}, \au{Pieters, G.J.M.}, \au{Wooding, R.A.} \&
  \au{van~der Ploeg, A.}} \yr{2002}  \at{Stability criteria for the vertical
  boundary layer formed by throughflow near the surface of a porous medium}.
  \bt{In {\em Environmental Mechanics: Water, Mass and Energy Transfer in the
  Biosphere\/} (ed. \ed{Peter~A.C. Raats, David Smiles \& Arthur~W. Warrick})},
   \pg{pp. 155--169}.  \publ{American Geophysical Union}.

\bibitem[Elder(1967)]{ELDER1967}
{\sc \au{Elder, J.W.}} \yr{1967}  \at{Steady free convection in a porous medium
  heated from below}.  \jt{J. Fluid Mech.}  \bvol{27}~(1),  \pg{29--48}.

\bibitem[Eloukabi {\em et~al.\/}(2013)Eloukabi, Sghaier, Nasrallah \&
  Prat]{Eloukabi2013}
{\sc \au{Eloukabi, H.}, \au{Sghaier, N.}, \au{Nasrallah, S.~Ben} \& \au{Prat,
  M.}} \yr{2013}  \at{{Experimental study of the effect of sodium chloride on
  drying of porous media: the crusty–patchy efflorescence transition}}.
  \jt{Int. J. Heat Mass Trans.}  \bvol{56},  \pg{80--93}.

\bibitem[Ernst(2017)]{ernst2017numerical}
{\sc \au{Ernst, Marcel}} \yr{2017} Numerical simulation of polygonal patterns
  in salt playa. Master's thesis, Georg-August-Universit{\"a}t G{\"o}ttingen.

\bibitem[Farhat(2018)]{Farhat2018}
{\sc \au{Farhat, Nasser}} \yr{2018}  \at{{Effect of relative humidity on
  evaporation rates in Nabatieh region}}.  \jt{Leban. Sci. J.}  \bvol{19},
  \pg{59--66}.

\bibitem[Fu {\em et~al.\/}(2014)Fu, Cueto-Felgueroso \& Juanes]{FU2014}
{\sc \au{Fu, Xiaojing}, \au{Cueto-Felgueroso, Luis} \& \au{Juanes, Ruben}}
  \yr{2014}  \at{Pattern formation and coarsening dynamics in three-dimensional
  convective mixing in porous media}.  \jt{Phil. Trans. R. Soc. A}  \bvol{371},
   \pg{20120355}.

\bibitem[Gill(1996)]{GILL1996}
{\sc \au{Gill, Thomas~E.}} \yr{1996}  \at{{Eolian sediments generated by
  anthropogenic disturbance of playas: Human impacts on the geomorphic system
  and geomorphic impacts on the human system}}.  \jt{Geomorphology}
  \bvol{17}~(1),  \pg{207--228}.

\bibitem[Graham \& Steen(1994)]{GRAHAM1994}
{\sc \au{Graham, Michael~D.} \& \au{Steen, Paul~H.}} \yr{1994}  \at{Plume
  formation and resonant bifurcations in porous-media convection}.  \jt{J.
  Fluid Mech.}  \bvol{272},  \pg{67–--90}.

\bibitem[Groeneveld {\em et~al.\/}(2010)Groeneveld, Huntington \&
  Barz]{GROENEVELD2010}
{\sc \au{Groeneveld, D.P.}, \au{Huntington, J.L.} \& \au{Barz, D.D.}} \yr{2010}
   \at{{Floating brine crusts, reduction of evaporation and possible
  replacement of fresh water to control dust from Owens Lake bed, California}}.
   \jt{J. Hydrol.}  \bvol{392}~(3),  \pg{211--218}.

\bibitem[Harfash(2013)]{HARFASH2013}
{\sc \au{Harfash, A.J.}} \yr{2013}  \at{Three-dimensional simulations for
  convection problem in anisotropic porous media with nonhomogeneous porosity,
  thermal diffusivity, and variable gravity effects}.  \jt{Transport Porous
  Med.}  \bvol{102}~(1),  \pg{43--57}.

\bibitem[Hewitt(2020)]{HEWITT2020b}
{\sc \au{Hewitt, D.R.}} \yr{2020}  \at{Vigorous convection in porous media}.
  \jt{Proc. R. Soc. A}  \bvol{476},  \pg{20200111}.

\bibitem[Hewitt {\em et~al.\/}(2012)Hewitt, Neufeld \& Lister]{HEWITT2012}
{\sc \au{Hewitt, Duncan~R.}, \au{Neufeld, Jerome~A.} \& \au{Lister, John~R.}}
  \yr{2012}  \at{{Ultimate regime of high Rayleigh number convection in a
  porous medium}}.  \jt{Phys. Rev. Lett.}  \bvol{108},  \pg{224503}.

\bibitem[Hewitt {\em et~al.\/}(2014)Hewitt, Neufeld \& Lister]{HEWITT2014}
{\sc \au{Hewitt, Duncan~R.}, \au{Neufeld, Jerome~A.} \& \au{Lister, John~R.}}
  \yr{2014}  \at{{High Rayleigh number convection in a porous medium containing
  a thin low-permeability layer}}.  \jt{J. Fluid Mech.}  \bvol{756},
  \pg{844--869}.

\bibitem[Hewitt {\em et~al.\/}(2020)Hewitt, Peng \& Lister]{HEWITT2020}
{\sc \au{Hewitt, Duncan~R.}, \au{Peng, Gunnar~G.} \& \au{Lister, John~R.}}
  \yr{2020}  \at{Buoyancy-driven plumes in a layered porous medium}.  \jt{J.
  Fluid Mech.}  \bvol{883},  \pg{A37}.

\bibitem[Homsy \& Sherwood(1976)]{HOMSY1976}
{\sc \au{Homsy, George~M.} \& \au{Sherwood, Albert~E.}} \yr{1976}
  \at{Convective instabilities in porous media with through flow}.  \jt{AIChE
  J.}  \bvol{22}~(1),  \pg{168--174}.

\bibitem[Horton \& Rogers(1945)]{HORTON1945}
{\sc \au{Horton, C.W.} \& \au{Rogers, F.T.}} \yr{1945}  \at{Convection currents
  in a porous medium}.  \jt{J. Appl. Phys.}  \bvol{16}~(6),  \pg{367--370}.

\bibitem[Koekoek \& Swarttouw(1998)]{KOEKOEK1998}
{\sc \au{Koekoek, Roelof} \& \au{Swarttouw, Ren\'e~F.}} \yr{1998} {The
  Askey-scheme of hypergeometric orthogonal polynomials and its q-analogue}.
  Delft University of Technology, Faculty of Information Technology and
  Systems, Department of Technical Mathematics and Informatics, Report no.
  98-17.

\bibitem[Krinsley(1970)]{KRINSLEY1970}
{\sc \au{Krinsley, Daniel~B.}} \yr{1970} {A geomorphological and
  paleoclimatological study of the playas of {Iran}. {Part} 1}. U.S. Geological
  Survey, final scientific report, contract no. PRO CP 70-800.

\bibitem[Lapwood(1948)]{LAPWOOD1948}
{\sc \au{Lapwood, E.R.}} \yr{1948}  \at{Convection of a fluid in a porous
  medium}.  \jt{Math. Proc. Cambridge}  \bvol{44}~(4),  \pg{508–--521}.

\bibitem[Lasser(2019)]{LASSER_DISS2019}
{\sc \au{Lasser, Jana}} \yr{2019}  \at{Geophysical pattern formation of salt
  playa}. PhD thesis, Georg-August-Universit{\"a}t G\"ottingen.

\bibitem[Lasser \& Ernst(2020)]{simulationcode}
{\sc \au{Lasser, Jana} \& \au{Ernst, Marcel}} \yr{2020}
  salt-playa-convection-simulation. DOI: 10.5281/zenodo.3969492.

\bibitem[{Lasser} \& {Goehring}(2020)]{CONCENTRATION}
{\sc \au{{Lasser}, Jana} \& \au{{Goehring}, Lucas}} \yr{2020} {Subsurface salt
  concentration profiles and pore water density measurements from Owens Lake,
  central California, measured in 2018}. {PANGAEA,
  https://doi.org/10.1594/PANGAEA.911059}.

\bibitem[Lasser {\em et~al.\/}(2019)Lasser, Nield, Ernst, Karius, Wiggs \&
  Goehring]{LASSER2019}
{\sc \au{Lasser, J.}, \au{Nield, J.M.}, \au{Ernst, M.}, \au{Karius, V.},
  \au{Wiggs, G.F.S.} \& \au{Goehring, L.}} \yr{2019} Salt polygons are caused
  by convection. ArXiv:1902.03600v2 [nlin.PS].

\bibitem[Lasser {\em et~al.\/}(2020)Lasser, Nield \&
  Goehring]{lasser2020surface}
{\sc \au{Lasser, Jana}, \au{Nield, Joanna~M.} \& \au{Goehring, Lucas}}
  \yr{2020}  \at{Surface and subsurface characterisation of salt pans
  expressing polygonal patterns}.  \jt{Earth Syst. Sci. Dat}  \bvol{12}~(4),
  \pg{2881--2898}.

\bibitem[Lele(1992)]{LELE1992}
{\sc \au{Lele, Sanjiva~K.}} \yr{1992}  \at{Compact finite difference schemes
  with spectral-like resolution}.  \jt{J. Comput. Phys.}  \bvol{103}~(1),
  \pg{16--42}.

\bibitem[Li {\em et~al.\/}(2019)Li, Cai, Li, Li \& Chen]{LI2019}
{\sc \au{Li, Qian}, \au{Cai, Weihua}, \au{Li, Feng-Chen}, \au{Li, Bingxi} \&
  \au{Chen, Ching-Yao}} \yr{2019}  \at{Miscible density-driven flows in
  heterogeneous porous media: Influences of correlation length and distribution
  of permeability}.  \jt{Phys. Rev. Fluids}  \bvol{4}~(1),  \pg{014502}.

\bibitem[Liang {\em et~al.\/}(2018)Liang, Wen, Hesse \& DiCarlo]{Liang2018}
{\sc \au{Liang, Yu}, \au{Wen, Baole}, \au{Hesse, Marc~A.} \& \au{DiCarlo,
  David}} \yr{2018}  \at{Effect of dispersion on solutal convection in porous
  media}.  \jt{Geophys. Res. Lett.}  \bvol{45},  \pg{9690--9698}.

\bibitem[Lokier(2012)]{LOKIER2012}
{\sc \au{Lokier, S.W.}} \yr{2012}  \at{{Development and evolution of subaerial
  halite crust morphologies in a coastal Sabkha setting}}.  \jt{J. Arid
  Environ.}  \bvol{79},  \pg{32 -- 47}.

\bibitem[Loodts {\em et~al.\/}(2014)Loodts, Rongy \& De~Wit]{LOODTS2014}
{\sc \au{Loodts, V.}, \au{Rongy, L.} \& \au{De~Wit, A.}} \yr{2014}  \at{Impact
  of pressure, salt concentration, and temperature on the convective
  dissolution of carbon dioxide in aqueous solutions}.  \jt{Chaos}
  \bvol{24}~(4),  \pg{043120}.

\bibitem[Lowenstein \& Hardie(1985)]{LOWENSTEIN1985}
{\sc \au{Lowenstein, Tim~K.} \& \au{Hardie, Lawrence~A.}} \yr{1985}
  \at{Criteria for the recognition of salt-pan evaporites}.  \jt{Sedimentology}
   \bvol{32}~(5),  \pg{627--644}.

\bibitem[Marticorena \& Bergametti(1995)]{MARTICORENA1995}
{\sc \au{Marticorena, B.} \& \au{Bergametti, G.}} \yr{1995}  \at{{Modeling the
  atmospheric dust cycle: 1. Design of a soil-derived dust emission scheme}}.
  \jt{J. Geophys. Res.}  \bvol{100}~(D8),  \pg{16415--16430}.

\bibitem[Metz {\em et~al.\/}(2005)Metz, Davidson, Coninck, Loos \&
  Meyer]{METZ2005}
{\sc \au{Metz, Bert}, \au{Davidson, Ogunlade}, \au{Coninck, Heleen~de},
  \au{Loos, Manuela} \& \au{Meyer, Leo}}, ed. \yr{2005} {\em {IPCC Special
  Report on Carbon Dioxide Capture and Storage. Prepared by Working Group III
  of the Intergovernmental Panel on Climate Change}\/}.  \publ{Cambridge
  University Press}.

\bibitem[Mojtabi \& Charrier-Mojtabi(2005)]{Mojtabi2005}
{\sc \au{Mojtabi, Abdelkader} \& \au{Charrier-Mojtabi, Marie-Catherine}}
  \yr{2005}  \at{{Double-diffusive convection in porous media}}.  \bt{In {\em
  Handbook of Porous Media\/} (ed. \ed{Kambiz Vafai})},  \pg{pp. 269--320}.
  \publ{CRC Press}.

\bibitem[Nachshon {\em et~al.\/}(2018)Nachshon, Weisbrod, Katzir \&
  Nasser]{Nachshon2018}
{\sc \au{Nachshon, Uri}, \au{Weisbrod, Noam}, \au{Katzir, Roee} \& \au{Nasser,
  Ahmed}} \yr{2018}  \at{{NaCl crust architecture and its impact on
  evaporation: three-dimensional insights}}.  \jt{Geophys. Res. Lett.}
  \bvol{45},  \pg{6100--6108}.

\bibitem[Neufeld {\em et~al.\/}(2010)Neufeld, Hesse, Riaz, Hallworth, Tchelepi
  \& Huppert]{NEUFELD2010}
{\sc \au{Neufeld, Jerome~A.}, \au{Hesse, Marc~A.}, \au{Riaz, Amir},
  \au{Hallworth, Mark~A.}, \au{Tchelepi, Hamdi~A.} \& \au{Huppert, Herbert~E.}}
  \yr{2010}  \at{Convective dissolution of carbon dioxide in saline aquifers}.
  \jt{Geophys. Res. Lett.}  \bvol{37}~(22),  \pg{L22404}.

\bibitem[Nield {\em et~al.\/}(2015)Nield, Bryant, Wiggs, King, Thomas, Eckardt
  \& Washington]{NIELD2015}
{\sc \au{Nield, J.M.}, \au{Bryant, R.G.}, \au{Wiggs, G.F.S.}, \au{King, J.},
  \au{Thomas, D.S.G.}, \au{Eckardt, F.D.} \& \au{Washington, R.}} \yr{2015}
  \at{The dynamism of salt crust patterns on playas}.  \jt{Geology}
  \bvol{43}~(1),  \pg{31--34}.

\bibitem[{Nield} {\em et~al.\/}(2020){Nield}, {Lasser} \& {Goehring}]{TEMPHUM}
{\sc \au{{Nield}, Joanna~M.}, \au{{Lasser}, Jana} \& \au{{Goehring}, Lucas}}
  \yr{2020} {Temperature and humidity time-series from Owens Lake, central
  California, measured during one week in November 2016}. {PANGAEA,
  https://doi.org/10.1594/PANGAEA.911059}.

\bibitem[Nield {\em et~al.\/}(2016)Nield, Neuman, O'Brien, Bryant \&
  Wiggs]{Nield2016}
{\sc \au{Nield, Joanna~M.}, \au{Neuman, Cheryl~McKenna}, \au{O'Brien, Patrick},
  \au{Bryant, Robert~G.} \& \au{Wiggs, Giles~F.S.}} \yr{2016}  \at{{Evaporative
  sodium salt crust development and its wind tunnel derived transport dynamics
  under variable climatic conditions}}.  \jt{Aeolian Res.}  \bvol{23},
  \pg{51--62}.

\bibitem[Otero {\em et~al.\/}(2004)Otero, Dontcheva, Johnston, Worthing,
  Kurganov, Petrova \& Doering]{OTERO2004}
{\sc \au{Otero, Jesse}, \au{Dontcheva, Lubomira~A.}, \au{Johnston, Hans},
  \au{Worthing, Rodney~A.}, \au{Kurganov, Alexander}, \au{Petrova, Guergana} \&
  \au{Doering, Charles~R.}} \yr{2004}  \at{{High-Rayleigh-number convection in
  a fluid-saturated porous layer}}.  \jt{J. Fluid Mech.}  \bvol{500},
  \pg{263–--281}.

\bibitem[Pellew \& Southwell(1940)]{pellew1940maintained}
{\sc \au{Pellew, Anne} \& \au{Southwell, R.V, .}} \yr{1940}  \at{On maintained
  convective motion in a fluid heated from below}.  \jt{Proc. R. Soc. A}
  \bvol{176}~(966),  \pg{312--343}.

\bibitem[Photographersnature(2012)]{PHOTOGRAPHERSNATURE}
{\sc \au{Photographersnature}} \yr{2012} {Death Valley's Badwater salt flats at
  twilight}.
  \url{https://en.wikipedia.org/wiki/Badwater_Basin#/media/File:Badwater_Salt_Flats_at_Twilight.jpg},
  acceses 2020-01-03.

\bibitem[Prospero(2002)]{PROSPERO2002}
{\sc \au{Prospero, J.M.}} \yr{2002}  \at{Environmental characterization of
  global sources of atmospheric soil dust identified with the {NIMBUS} 7 total
  ozone mapping spectrometer ({TOMS}) absorbing aerosol product}.  \jt{Rev.
  Geophys.}  \bvol{40}~(1),  \pg{2--1--2--31}.

\bibitem[Rapaka {\em et~al.\/}(2008)Rapaka, Chen, Pawar, Stauffer \&
  Zhang]{RAPAKA2008}
{\sc \au{Rapaka, Saikiran}, \au{Chen, Shiyi}, \au{Pawar, Rajesh~J.},
  \au{Stauffer, Philip~H.} \& \au{Zhang, Dongxiao}} \yr{2008}  \at{Non-modal
  growth of perturbations in density-driven convection in porous media}.
  \jt{J. Fluid Mech.}  \bvol{609},  \pg{285--303}.

\bibitem[Raupach {\em et~al.\/}(1993)Raupach, Gillette \& Leys]{RAUPACH1993}
{\sc \au{Raupach, M.R.}, \au{Gillette, D.A.} \& \au{Leys, J.F.}} \yr{1993}
  \at{The effect of roughness elements on wind erosion threshold}.  \jt{J.
  Geophys. Res. Atmos.}  \bvol{98}~(D2),  \pg{3023--3029}.

\bibitem[Riaz \& Meiburg(2003)]{RIAZ2003}
{\sc \au{Riaz, A.} \& \au{Meiburg, E.}} \yr{2003}  \at{Three-dimensional
  miscible displacement simulations in homogeneous porous media with gravity
  override}.  \jt{J. Fluid Mech.}  \bvol{494},  \pg{95--117}.

\bibitem[Ruith \& Meiburg(2000)]{RUITH2000}
{\sc \au{Ruith, Michael} \& \au{Meiburg, Eckart}} \yr{2000}  \at{{Miscible
  rectilinear displacements with gravity override. Part 1. Homogeneous porous
  medium}}.  \jt{J. Fluid Mech.}  \bvol{420},  \pg{225--257}.

\bibitem[Sharp \& Shi(2009)]{SHARP2009}
{\sc \au{Sharp, J.M.} \& \au{Shi, M.}} \yr{2009}  \at{Heterogeneity effects on
  possible salinity-driven free convection in low-permeability strata}.
  \jt{Geofluids}  \bvol{9}~(4),  \pg{263--274}.

\bibitem[Slim(2014)]{SLIM}
{\sc \au{Slim, Anja~C.}} \yr{2014}  \at{Solutal-convection regimes in a
  two-dimensional porous medium}.  \jt{J. Fluid Mech.}  \bvol{741},
  \pg{461–--491}.

\bibitem[Slim {\em et~al.\/}(2013)Slim, Bandi, Miller \& Mahadevan]{SLIM2013}
{\sc \au{Slim, Anja~C.}, \au{Bandi, M.M.}, \au{Miller, Joel~C.} \&
  \au{Mahadevan, L.}} \yr{2013}  \at{{Dissolution-driven convection in a
  Hele-Shaw cell}}.  \jt{Phys. Fluids}  \bvol{25}~(2),  \pg{024101}.

\bibitem[Slim \& Ramakrishnan(2010)]{SLIM2010}
{\sc \au{Slim, Anja~C.} \& \au{Ramakrishnan, T.~S.}} \yr{2010}  \at{{Onset and
  cessation of time-dependent, dissolution-driven convection in porous media}}.
   \jt{Phys. Fluids}  \bvol{22}~(12),  \pg{124103}.

\bibitem[Stevens {\em et~al.\/}(2009)Stevens, Jr., Simmons \&
  Fenstemaker]{STEVENS2009}
{\sc \au{Stevens, Joel~D.}, \au{Jr., John M.~Sharp}, \au{Simmons, Craig~T.} \&
  \au{Fenstemaker, T.R.}} \yr{2009}  \at{Evidence of free convection in
  groundwater: Field-based measurements beneath wind-tidal flats}.  \jt{J.
  Hydrol.}  \bvol{375},  \pg{394--409}.

\bibitem[S\"uli(2003)]{SULI2003}
{\sc \au{S\"uli, Endre}} \yr{2003} {\em An Introduction to Numerical
  Analysis\/}.  \publ{Cambridge University Press}.

\bibitem[Tan \& Homsy(1988)]{TAN1988}
{\sc \au{Tan, C.T.} \& \au{Homsy, G.M.}} \yr{1988}  \at{Simulation of nonlinear
  viscous fingering in miscible displacement}.  \jt{Phys. Fluids}
  \bvol{31}~(6),  \pg{1330--1338}.

\bibitem[Taylor(1953)]{TAYLOR1953}
{\sc \au{Taylor, Geoffrey}} \yr{1953}  \at{Dispersion of soluble matter in
  solvent flowing slowly through a tube}.  \jt{Proc. R. Soc. A}
  \bvol{219}~(1137),  \pg{186--203}.

\bibitem[Thomas {\em et~al.\/}(2018)Thomas, Dehaeck \& {De Wit}]{THOMAS2018}
{\sc \au{Thomas, C.}, \au{Dehaeck, S.} \& \au{{De Wit}, A.}} \yr{2018}
  \at{{Convective dissolution of CO$_2$ in water and salt solutions}}.
  \jt{Int. J. Greenh. Gas Con.}  \bvol{72},  \pg{105--116}.

\bibitem[Trefethen(2000)]{TREFETHEN2000}
{\sc \au{Trefethen, Lloyd~N.}} \yr{2000} {\em Spectral Methods in MATLAB\/}.
  \publ{Society for Industrial and Applied Mathematics}.

\bibitem[Tyler(2007)]{TYLER2007}
{\sc \au{Tyler, Jonathan~G.}} \yr{2007}  \at{Analysis and implementation of
  high-order compact finite difference schemes}. PhD thesis, Brigham Young
  University.

\bibitem[Tyler {\em et~al.\/}(1997)Tyler, Kranz, Parlange, Albertson, Katul,
  Cochran, Lyles \& Holder]{TYLER1997}
{\sc \au{Tyler, S.W.}, \au{Kranz, S.}, \au{Parlange, M.B.}, \au{Albertson, J.},
  \au{Katul, G.G.}, \au{Cochran, G.F.}, \au{Lyles, B.A.} \& \au{Holder, G.}}
  \yr{1997}  \at{{Estimation of groundwater evaporation and salt flux from
  Owens Lake, California, USA}}.  \jt{J. Hydrol.}  \bvol{200}~(1-4),
  \pg{110--135}.

\bibitem[Van~Dam {\em et~al.\/}(2009)Van~Dam, Simmons, Hyndman \&
  Wood]{VANDAM2009}
{\sc \au{Van~Dam, Remke~L.}, \au{Simmons, Craig~T.}, \au{Hyndman, David~W.} \&
  \au{Wood, Warren~W.}} \yr{2009}  \at{Natural free convection in porous media:
  First field documentation in groundwater}.  \jt{Geophys. Res. Lett.}
  \bvol{36}~(11),  \pg{L11403}.

\bibitem[Washington {\em et~al.\/}(2003)Washington, Todd, Middleton \&
  Goudie]{WASHINGTON2003}
{\sc \au{Washington, R.}, \au{Todd, M.}, \au{Middleton, N.J.} \& \au{Goudie,
  A.S.}} \yr{2003}  \at{Dust-storm source areas determined by the total ozone
  monitoring spectrometer and surface observations}.  \jt{Ann. Assoc. Am.
  Geogr.}  \bvol{93}~(2),  \pg{297--313}.

\bibitem[Wooding(1960)]{WOODING1960a}
{\sc \au{Wooding, R.A.}} \yr{1960}  \at{Rayleigh instability of a thermal
  boundary layer in flow through a porous medium}.  \jt{J. Fluid Mech.}
  \bvol{9}~(2),  \pg{183--192}.

\bibitem[Wooding {\em et~al.\/}(1997)Wooding, Tyler \& White]{WOODING1997a}
{\sc \au{Wooding, R.A.}, \au{Tyler, Scott~W.} \& \au{White, Ian}} \yr{1997}
  \at{{Convection in groundwater below an evaporating salt lake: 1. Onset of
  instability}}.  \jt{Water Resour. Res.}  \bvol{33}~(6),  \pg{1199--1217}.

\end{thebibliography}

\end{document}